\documentclass[reprint]{revtex4-2}
\usepackage{geometry}
\usepackage{physics}
\usepackage[version=4]{mhchem}
\usepackage{amsmath}
\usepackage[super]{nth}
\usepackage{amsfonts}
\usepackage{siunitx}
\usepackage{float}
\usepackage[caption=false]{subfig}
\usepackage{rotating}
\usepackage{hyperref}
    \hypersetup{colorlinks=true,
                linkcolor=red,
                filecolor=blue,
                citecolor=blue,      
                urlcolor=cyan,
    }
\usepackage{bm}
\usepackage{titlesec}
    \titlespacing\section{0pt}{12pt plus 4pt minus 2pt}{0pt plus 2pt minus 2pt}
    \titlespacing\subsection{0pt}{12pt plus 4pt minus 2pt}{0pt plus 2pt minus 2pt}
    \titlespacing\subsubsection{0pt}{12pt plus 4pt minus 2pt}{0pt plus 2pt minus 2pt}
\usepackage{natbib}

\begin{document}

\title{\textbf{A Quantum Simulation Approach to Implementing Nuclear Density Functional Theory via Imaginary Time Evolution}}
\author{Yang Hong Li}
\author{Jim Al-Khalili}
\author{Paul Stevenson}
\affiliation{School of Mathematics and Physics, Faculty of Engineering and Physical Sciences \\University of Surrey, Guildford, GU2 7XH, UK}
\date{\today}

\begin{abstract}
The quantum imaginary time evolution (QITE) algorithm is a direct implementation of the classical imaginary time evolution algorithm on quantum computer.  We implement the QITE algorithm for the case of nuclear Hartree-Fock equations in a formalism equivalent to nuclear density functional theory.  We demonstrate the algorithm in the case of the helium-4 nucleus with a simplified effective interaction of the Skyrme kind and demonstrate that the QITE, as implemented on simulated quantum computer, gives identical results to the classical algorithm.  
\end{abstract}

\maketitle
\section{Introduction}
Quantum computers have developed from hypothetical devices to real systems capable of harnessing quantum entanglement and the efficient representation of information to bring a new paradigm to computation and information processing.  One major area of application of quantum computers is in the simulation of other quantum systems, where qubit encoding of wave functions of many-particle systems can be enacted particularly efficiently.  These ideas have been used in quantum fermionic systems such as quantum chemistry 
\cite{RevModPhys.92.015003,ChemRev.119.10856,Nature.618.500,Science.369.1084}, materials \cite{NatComputSci.2.424,QuantumSciTechnol.6.043002}, and nuclear physics \cite{TESNAT2015.01008,ChinPhysB.30.020306,IntJUnconvComput.18.83,AdvQuantumTechnol.2300219}.  In nuclear physics, general quantum algorithms such as the variational quantum eigensolver (VQE) and its extensions \cite{npjQuantumInf.8.1} have been widely used to find energy eigenvalues in systems such as the deuteron, as represented through an effective field theory \cite{PhysRevLett.120.210501,PhysRevC.104.034301}, the Lipkin Meshkov Glick model \cite{PhysRevA.104.022412,ChinPhysC.46.024106,PhysRevC.106.024319,PhysRevC.105.064317,SPIEPhotonicsEurope.109,PhysRevC.108.024313}, and the nuclear shell-model \cite{PhysRevC.105.064308,PhysRevC.106.034325,PhysRevC.105.064317,PhysRevC.108.064305,SciRep.13.12291}.   

In the present work, we are concerned with the development and implementation of a quantum computational algorithm of the imaginary time evolution type to solve the nuclear Hartree-Fock equations.  We use a simplified but typical nuclear effective interaction of the Skyrme form that gives rise to the usual non-linear Schr\"odinger equation, which can also be cast as a density functional theory problem.  Although Hartree-Fock solutions are achievable on classical computers, implementation on quantum computer is a kind of benchmark \cite{Science.369.1084}, and we intend that our algorithm can serve as a starting point for much more sophisticated models with highly-correlated states, where quantum algorithms should be more useful. The procedure can be considered as a kind of state preparation algorithm for deeper quantum simulation, itself a topic of general interest \cite{PhysRevLett.129.230504}, or as the starting point for quantum algorithms such as symmetry restoration as proposed for use with nuclear-like systems
\cite{PhysRevC.105.024324,PhysRevC.107.034310}.  We also note that the solution, on quantum computer, of the type of nonlinear Schr\"odinger equation presented here is of interest beyond nuclear physics \cite{PhysRevA.101.010301}.

In what follows, we give a brief summary of the nuclear physics problem, a description of the imaginary time evolution algorithm with our particular quantum implementation, and follow with results in the simplest case -- a spherical $^4$He nucleus in the absence of a Coulomb interaction and where there is a single ($s_{1/2}$) single-particle wave function to determine.

\section{Nuclear Model}
For a nucleus consisting of $N$ strongly interacting nucleons,  the time-independent Schr\"{o}dinger equation governing the $N$-body wavefunction  $\Phi(\{x_i\})$ is
\begin{equation}\label{NbodySE}
    \left[-\frac{\hbar^2}{2m}\sum_{i=1}^{N}\nabla_i^2+V(\{x_i\})\right]\Phi=\mathcal{E}\Phi\text{,}
\end{equation}
where $x=\left\{\vec{r}, S, I_3\right\}$ includes all of space, spin, and isospin coordinates. This equation cannot be solved exactly in general and reasonable approximations are usually required. One common method for describing the nuclear structure and low energy dynamics \cite{RevModPhys.54.913,JPhysG.46.013001} is the mean field approximation. By treating each of the $N$ interacting particles as a single particle in the field created by the remaining $N-1$ particles, the $N$-body problem is essentially reduced to a self-consistent one-body problem.  Accompanying the mean-field approximation is the form of the nuclear interaction, which needs to be appropriate for use in the mean-field limit; either in the form of a renormalized realistic interaction, or a phenomenological one.  We use here the latter form, of the Skyrme kind \cite{NuclPhys.9.615,PhysRevC.5.626}. 

The Skyrme interaction starts with a potential consisting of a two-body term and a three-body term 
\begin{equation}\begin{aligned}
    V(\{x_i\})&=\frac{1}{2}\sum_{i\neq j=1}^{N}v^{(2)}\left(x_i,x_j\right)\\
    &\quad+\frac{1}{6}\sum_{i\neq j\neq k=1}^{N}v^{(3)}\left(x_i,x_j,x_k\right)\text{.}
\end{aligned}\end{equation}

\noindent The three-body term is assumed to be a zero-range force
\begin{equation}
    v^{(3)}_{ijk}=t_3\delta^3\left(\vec{r}_i-\vec{r}_j\right)\delta^3\left(\vec{r}_j-\vec{r}_k\right)\text{,}
\end{equation}
while the two-body term is assumed to be short ranged, with matrix elements in momentum space given by
\begin{equation}\label{skyrme2}\begin{aligned}
    \mel{\vec{k}}{v^{(2)}_{ij}}{\vec{k}'}&=t_0\left(1+x_0P_\sigma\right)\\
    &\quad+\frac{1}{2}t_1\left(k^2+{k'}^2\right)+t_2\vec{k}\cdot\vec{k}'\\
    &\quad+iW_0\left(\vec{\sigma}_i+\vec{\sigma}_j\right)\cdot\vec{k}\times\vec{k}'\text{,}
\end{aligned}\end{equation}
where $P_\sigma$, $\vec{\sigma}$, and $\vec{k}$ are the spin-exchange operator, Pauli spin matrices, and relative wave vector respectively (denoting $v^{(2)}\left(x_i,x_j\right)$ as $v^{(2)}_{ij}$ and $v^{(3)}\left(x_i,x_j,x_k\right)$ as $v^{(3)}_{ijk}$ ). In configuration space equation (\ref{skyrme2}) becomes
\begin{equation}\begin{aligned}
    v^{(2)}_{ij}&=t_0\left(1+x_0P_\sigma\right)\delta^3\left(\vec{r}_i-\vec{r}_j\right)\\
    &\quad+\frac{1}{2}t_1\left[\delta^3\left(\vec{r}_i-\vec{r}_j\right)k^2_R+k^2_L\delta^3\left(\vec{r}_i-\vec{r}_j\right)\right]\\
    &\quad+t_2\vec{k}_L\cdot\delta^3\left(\vec{r}_i-\vec{r}_j\right)\vec{k}_R\\ &\quad+iW_0\left(\vec{\sigma}_i+\vec{\sigma}_j\right)\cdot\vec{k}_L\times\delta^3\left(\vec{r}_i-\vec{r}_j\right)\vec{k}_R\text{,}
\end{aligned}\end{equation}
where $\vec{k}_{L,R}=\frac{1}{2}i\left(\vec{\nabla}_i-\vec{\nabla}_j\right)$ are relative wave vector operators, with the subscripts $L,R$ denoting left-multiplying and right-multiplying respectively.

\noindent It can be shown that \cite{PhysRevC.5.626}, for Hartree-Fock calculations, in which the nuclear wave function is assumed to be a Slater determinant of single particle states, the three-body term is equivalent to the density-dependent two-body interaction
\begin{equation}
    v^{(3)}_{ij}=\frac{1}{6}t_3\left(1+P_\sigma\right)\delta^3\left(\vec{r}_i-\vec{r}_j\right)\rho\left(\frac{\vec{r}_i+\vec{r}_j}{2}\right)\text{,}
\end{equation}
where $\rho\left(\vec{r}\right)$ is the single-particle density.

\noindent Considering the case of a nucleus with $A=2Z$ (e.g. \ce{^{4}He}, \ce{^{16}O}), the further simplifications of the absence of a Coulomb field, and with a simplified Skyrme force in which the $t_1$, $t_2$, and $W_0$ terms are neglected, we have a simplified model often used for exploratory studies \cite{PhysRevC.60.044302,PhysRevC.87.014330}. The potential is then reduced to an effective two-body interaction
\begin{equation}\label{Skyrme_potential}\begin{aligned}
    v\left(x_i,x_j\right)&=v^{(2)}_{ij}+v^{(3)}_{ij}\\
    &=\frac{3}{8}t_0\delta^3\left(\vec{r}_i-\vec{r}_j\right)\\
    &\quad+\frac{1}{16}t_3\delta^3\left(\vec{r}_i-\vec{r}_j\right)\rho\left(\frac{\vec{r}_i+\vec{r}_j}{2}\right).
\end{aligned}\end{equation}

\noindent With a Slater determinant wave function, the expectation value of the Hamiltonian (i.e. the energy) is

\begin{equation}\begin{aligned}
    E&=-\frac{\hbar^2}{2m}\sum_{i=1}^{N}\int d^3\vec{r}\left[\varphi_i^*\left(\vec{r}\right)\nabla_i^2\varphi_i\left(\vec{r}\right)\right]\\
    &\quad+\int d^3\vec{r}\left[\frac{3}{8}t_0\rho^2\left(\vec{r}\right)+\frac{1}{16}t_3\rho^3\left(\vec{r}\right)\right]\text{.}
\end{aligned}\end{equation}

\noindent By minimizing $E$ via the variational principle, while requiring that the single particle (SP) states are normalized, 
\begin{equation}
    \frac{\delta}{\delta\varphi_i^*}\left[E-\sum_{j}\varepsilon_j\int\varphi_j^*\left(\vec{r}\right)\varphi_j\left(\vec{r}\right)d^3\vec{r}\right]=0\text{,}
\end{equation}
we obtain the Hartree-Fock equations \cite{PhysRevC.60.044302}
\begin{equation}\label{HF_eqt}
    \left[-\frac{\hbar^2}{2m}\nabla^2+\frac{3}{4}t_0\rho\left(\vec{r}\right)+\frac{3}{16}t_3\rho^2\left(\vec{r}\right)\right]\varphi_j\left(\vec{r}\right)=\varepsilon_j\varphi_j\left(\vec{r}\right).
\end{equation}
  The Lagrange multipliers $\varepsilon_j$ can be identified as the SP energies.

\noindent For spherically-symmetric nuclei in the absence of spin-orbit interaction, the SP wave functions $\varphi_{nlmm_sq}\left(\vec{r},s,I_3\right)$ can be factorised into radial, angular, spin, and isospin parts, in which the angular and spin parts remain uncoupled:
\begin{equation} \label{SepVar}
    \varphi_{nlmm_sq}\left(\vec{r},s,I_3\right)=\frac{u_{nl}\left(r\right)}{r}Y_{lm}\left(\theta,\phi\right)\chi_{m_s}\left(s\right)\chi_q\left(I_3\right)\text{.}
\end{equation}
The spherical harmonics $Y_{lm}\left(\theta,\phi\right)$ form an orthonormal basis and describe the angular behaviour of $\varphi_{nlmm_sq}\left(\vec{r},s,I_3\right)$, while the spinors $\chi_{m_s}\left(s\right)$ and $\chi_q\left(I_3\right)$ represent the spin and isospin dependencies of $\varphi_{nlmm_sq}\left(\vec{r},s,I_3\right)$ respectively.

\noindent Equation (\ref{HF_eqt}) is then reduced to the radial equation
\begin{equation}\label{rad_eqt}
    \hat{H}_{\mathrm{HF}}^{l}u_{nl}(r)=\varepsilon_{nl}u_{nl}(r)\text{,}
\end{equation}
where
\begin{equation}\begin{aligned}
    \hat{H}_{\mathrm{hf}}^{l}&=-\frac{\hbar^2}{2m}\left[\frac{d^2}{dr^2}-\frac{l\left(l+1\right)}{r^2}\right]\\
    &\quad+\left[\frac{3}{4}t_0\rho\left(r\right)+\frac{3}{16}t_3\rho^2\left(r\right)\right]\text{,}
\end{aligned}\end{equation}
is the Hartree-Fock Hamiltonian. The density $\rho\left(r\right)$ is now given by
\begin{equation}\label{density}\begin{aligned}
    \rho\left(r\right)&=\frac{4}{r^2}\sum_{n,l}\left|u_{nl}\right|^2\sum^{l}_{m=-l}\left|Y_{lm}\right|^2\\
    &=\frac{1}{\pi r^2}\sum_{n,l}\left(2l+1\right)\left|u_{nl}\right|^2 \text{,}
\end{aligned}\end{equation}
with the factor of 4 in the first line arising from spin and isospin degeneracies.

\section{Imaginary Time Evolution}
In order to solve the nonlinear Schr\"{o}dinger equation (\ref{HF_eqt}), an iterative method is usually employed.  Here, we make use of the imaginary time evolution method.

The time-dependent Schr\"{o}dinger equation (TDSE) of the Hartree-Fock Hamiltonian $\hat{H}_{\mathrm{HF}}^{l}$
\begin{equation}
    \hat{H}_{hf}^{l}\ket{\psi\left(r,t\right)}=i\hbar\frac{\partial}{\partial t}\ket{\psi\left(r,t\right)}
\end{equation}
has the formal solution
\begin{equation}\label{time_ev}
    \ket{\psi\left(r,t\right)}=\exp(-\frac{i\hat{H}_{hf}^{l}t}{\hbar})\ket{\psi\left(r,0\right)}\text{.}
\end{equation}
Under imaginary time ($t\rightarrow -i\tau$), equation (\ref{time_ev}) becomes
\begin{equation}\label{im_time}
    \ket{\psi\left(r,\tau\right)}=\mathcal{N}\exp(-\frac{\hat{H}_{hf}^{l}}{\hbar}\tau)\ket{\psi\left(r,0\right)}\text{,}
\end{equation}
where $\mathcal{N}$ is a normalization operator to renormalize the state after the application of the non-unitary imaginary time evolution operator.
When $\tau\rightarrow\infty$, $\ket{\psi\left(r,\tau\right)}$ converges to $\ket{u_{0l}}$, the ground state of $\hat{H}_{\mathrm{HF}}^{l}$ provided that the initial state $\ket{\psi\left(r,0\right)}$ is not orthogonal to it. \cite{JComputPhys.221.148}.

\noindent In principle, the imaginary time evolution (ITE) operator $\hat{U}\left(\tau\right)=\exp(-\hat{H}_{\mathrm{HF}}^{l}\tau/\hbar)$ could be applied once, with a large enough imaginary time $\tau$, for a good approximation of the ground state. However, the true form of $\hat{H}_{\mathrm{HF}}^{l}$ is unknown due to its density dependence, and the ITE has to be separated into $k_{total}$ steps, each with an imaginary time step of $\Delta\tau=\frac{\tau}{k_{total}}$. This is equivalent to writing equation (\ref{im_time}) as  
\begin{equation}\label{im_time_ev}
    \ket{\psi\left(r,\tau\right)}=\left[\hat{U}\left(\Delta\tau\right)\right]^{k_{total}}\ket{\psi\left(r,0\right)}\text{,}
\end{equation}
where $\hat{U}$ is updated after each step using a newly obtained $\ket{\psi\left(r\right)}$. Providing the imaginary time step $\Delta\tau\ll1$ is small enough, $\hat{U}$ can be approximated by
\begin{equation}\label{eq:uapprox}
    \hat{U}\left(\Delta\tau\right)\approx1-\frac{\Delta\tau}{\hbar}\hat{H}_{\mathrm{HF}}^{l}\text{.}
\end{equation}

\subsection{Quantum Imaginary Time Algorithm}
There have been multiple attempts at implementing ITE on quantum devices \cite{PhysRevA.104.042418,PhysRevA.109.052414,arXiv:2306.14993,Quantum.7.916,PhysRevA.106.062435}. One of the main obstacles is the non-unitarity of the operator $\hat{U}$ as quantum circuits can only handle unitary operators. In the original QITE \cite{NatPhys.16.205} algorithm, this is overcome by sectioning the qubit chain and performing approximated Trotter evolution \cite{PacJMath.8.887,ProcAmMathSoc.10.545,CommunMathPhys.51.183} section by section. A later proposal \cite{JChemTheoryComput.19.3868} makes use of a randomized qDRIFT algorithm \cite{PhysRevLett.123.070503} which significantly reduces the circuit depth. In this paper, we use the idea of a duality computer \cite{CommunTheorPhys.45.825,Res.2020.1} to implement non-unitary Hermitian gates, with the aid of some ancillary qubits.

\noindent In order to encode the unknown wave function solution to the HF equation, we have chosen the 3D isotropic oscillator basis (with oscillator length $\frac{1}{b}$) \cite{PhysRev.177.1519,ComputPhysCommun.200.220,SciChinaPhysMechAstron.66.240311,ComputPhysCommun.102.166}
\begin{equation}\label{exp}
    u_{nl}\left(r\right)=\sum_{n'}\alpha_{nln'}\mathcal{R}^{b, l+\frac{3}{2}}_{n'}\left(r\right),\;\; \alpha_{nln'}\in\mathbb{R}\:\forall\:n,l,n'\text{,}
\end{equation} 
as our computational basis, where $\alpha_{nln'}$ are the expansion coefficients and $\mathcal{R}^{b, l+\frac{3}{2}}_{n'}\left(r\right)$ are the oscillator radial wavefunctions. In this basis, the matrix elements of the density are given by the sum of integrals of the product of four basis wavefunctions,
\begin{equation}
    \rho\sim\sum\int\frac{dr}{r^2}\mathcal{R}\mathcal{R}\mathcal{R}\mathcal{R}\text{.}
\end{equation}
These integrals are computed and tabulated classically at the beginning of the calculation, for evaluation of the density at each step of ITE, and a matrix representation of the time-evolution operator (\ref{eq:uapprox}) is constructed.

\noindent Using $N$ qubits, the first $n'=2^N$ coefficients of expansion $\alpha_{0ln'}$ can be represented as a state vector $\ket{\psi_{l}}$ such that
\begin{equation}\label{decomstate}
    \ket{{\psi_{l}}}=\sum_{n'=0}^{2^{N}-1}\alpha_{0ln'}\ket{n'}\text{,}
\end{equation}
where $\ket{0}=\ket{00\ldots00}$, $\ket{1}=\ket{00\ldots01}$, $\ket{2}=\ket{00\ldots10}$ \textit{etc.}

\noindent The ITE operator $\hat{U}$, previously obtained classically, is a $2^N\times2^N$ Hermitian matrix and can be decomposed into a sum of products of Pauli matrices (and identity matrices) acting on individual qubits \cite{arXiv:2111.00627}
\begin{align}\label{decomH}
    \hat{U}&=\sum_{i'=0}^{2^{2N}-1}\beta_{i'}P_{i'}\\
    P_{i'}&=\bigotimes_{q=1}^N\sigma_{i'_4\left[q\right]}\text{,}
\end{align}
where $i'_4\left[q\right]$ is the $q\textsuperscript{th}$ digit from the right of $i'$ when expressed in quaternary and
\begin{equation}
    \sigma_0=I,\;\;\sigma_1=Z,\;\;\sigma_2=X,\;\;\sigma_3=Y
\end{equation}
are the identity and Pauli matrices acting on the $\left(q-1\right)\textsuperscript{th}$ qubit. Using $i'=45$ as an example, since $45_{10}=231_{4}$, the corresponding gate is $X_2\otimes Y_1\otimes Z_0$. The coefficients $\beta_{i'}$ can then be stored in an ancillary state $\ket{\phi_a}$ using $2N$ ancillary qubits such that
\begin{equation}
    \ket{\phi_a}=\frac{1}{B}\sum_{i'=0}^{2^{2N}-1}\beta_{i'}\ket{i'}\text{,}
\end{equation}
where $B=\sqrt{\sum_{i'=0}^{2^{2N}-1}\beta_{i'}^2}$.

\noindent In the larger Hilbert space spanned by $\ket{\phi_a}$and $\ket{\psi_l}$
\begin{equation}
    \ket{\psi_l}\rightarrow\ket{\phi_a}\ket{\psi_l}=\frac{1}{B}\sum_{i'=0}^{2^{2N}-1}\beta_{i'}\ket{i'} \ket{\psi_l}\text{,}
\end{equation}
the outcome of the non-unitary $\hat{U}$ can be obtained in a particular subspace. This is achieved by applying a series of controlled Pauli gates
\begin{equation}\label{PauliOp}
    \hat{O}_P=\sum_{i'=0}^{2^{2N}-1}\ket{i'}\bra{i'}\otimes P_{i'}
\end{equation}
and $2N$ Hadamard gates
\begin{equation}
    \hat{O}_H=\bigotimes^{2N-1}_{q_a=0}H_{q_a}\text{.}
\end{equation}

\noindent The result of the operations is given by
\begin{equation}\label{HadComb}\begin{alignedat}{3}
    &\quad\hat{O}_H\hat{O}_P&&\ket{\phi_a}\ket{\psi_l}\\
    &=\frac{1}{2^NB}&&\left(\ket{0}\sum_{i'=0}^{2^{2N}-1}\beta_{i'}P_{i'}\ket{\psi_l}\right.\\
    &&&+\left.\ket{1}\sum_{i'=0}^{2^{2N}-1}\left(-1\right)^{i'}\beta_{i'}P_{i'}\ket{\psi_l}+\ldots\right)\\
    &=\frac{1}{2^NB}&&\left(\ket{0}\hat{U}\ket{\psi_l}+\ldots\right)\text{.}
\end{alignedat}\end{equation}
Thus we obtain our expected value and result state in the first $2^N$ entries of the statevector.

\subsection{$N=1$ case}
\noindent The details of the $N=1$ case help clarify the general algorithm.  We thus write out in full the $N=1$ case using three (one target and two ancillary) qubits, in which the ground state $\ket{\psi_l}$ is expanded in the two lowest oscillator states.

\noindent The ITE operator, $\hat{U}$, is then a $2\times2$ Hermitian matrix and its Pauli decomposition is
\begin{equation}
    \begin{aligned}
         \hat{U}=&\begin{pmatrix}U_{00}&U_{01}\\U_{10}&U_{11}\end{pmatrix}\\
         =&\frac{U_{00}+U_{11}}{2}I+\frac{U_{00}-U_{11}}{2}Z+\\
         &\frac{U_{01}+U_{10}}{2}X+i\frac{U_{01}-U_{10}}{2}Y.
    \end{aligned}
\end{equation}

\noindent Comparing with equation (\ref{decomH}), we see that
\begin{equation}\begin{aligned}
    \beta_0&=\frac{U_{00}+U_{11}}{2},\;\;\beta_1=\frac{U_{00}-U_{11}}{2},\\
    \beta_2&=\frac{U_{01}+U_{10}}{2},\;\;\beta_3=i\frac{U_{01}-U_{10}}{2},
    \end{aligned}
\end{equation}
and
\begin{equation}
    B=\sqrt{\frac{1}{2}\left(U^2_{00}+U^2_{11}\right)+U_{01}U_{10}}.
\end{equation}

\noindent Hence,
\begin{equation}\label{anstate}
    \ket{\phi_a}=\frac{1}{B}\left(\beta_0\ket{00}+\beta_1\ket{01}+\beta_2\ket{10}+\beta_3\ket{11}\right).
\end{equation}

\noindent For a general target state
\begin{equation}\label{target}
    \ket{\psi_l}=\begin{pmatrix}\alpha_{0l0}\\\alpha_{0l1}\end{pmatrix}\text{,}
\end{equation}
the expected output is
\begin{equation}\begin{aligned}\label{expect}
    \hat{U}\ket{\psi_l}&=\begin{pmatrix}U_{00}\alpha_{0l0}+U_{01}\alpha_{0l1}\\U_{10}\alpha_{0l0}+U_{11}\alpha_{0l1}\end{pmatrix}\\
    &=\begin{pmatrix}\left(\beta_0+\beta_1\right)\alpha_{0l0}+\left(\beta_2-i\beta_3\right)\alpha_{0l1}\\\left(\beta_2+i\beta_3\right)\alpha_{0l0}+\left(\beta_0-\beta_1\right)\alpha_{0l1}\end{pmatrix}\text{.}\end{aligned}
\end{equation}

\noindent Applying $\hat{O}_P$ from equation (\ref{PauliOp}), we obtain
\begin{equation}\begin{aligned}
    \hat{O}\ket{\phi_a}\ket{\psi_l}&=\begin{pmatrix}I&&&\\&Z&&\\&&X&\\&&&Y\end{pmatrix}\frac{1}{B}\begin{pmatrix}\beta_0\begin{pmatrix}\alpha_{0l0}\\\alpha_{0l1}\end{pmatrix}\\\beta_1\begin{pmatrix}\alpha_{0l0}\\\alpha_{0l1}\end{pmatrix}\\\beta_2\begin{pmatrix}\alpha_{0l0}\\\alpha_{0l1}\end{pmatrix}\\\beta_3\begin{pmatrix}\alpha_{0l0}\\\alpha_{0l1}\end{pmatrix}\end{pmatrix}\\
    &=\frac{1}{B}\begin{pmatrix}\beta_0\begin{pmatrix}\alpha_{0l0}\\\alpha_{0l1}\end{pmatrix}\\\beta_1\begin{pmatrix}\alpha_{0l0}\\-\alpha_{0l1}\end{pmatrix}\\\beta_2\begin{pmatrix}\alpha_{0l1}\\\alpha_{0l0}\end{pmatrix}\\i\beta_3\begin{pmatrix}-\alpha_{0l1}\\\alpha_{0l0}\end{pmatrix}\end{pmatrix}\text{.}\end{aligned}
\end{equation}

\noindent Combining the contribution using the Hadamard gates
\begin{equation}
    H_1\otimes H_2=\frac{1}{2}\begin{pmatrix}I&I&I&I\\I&-I&I&-I\\I&I&-I&-I\\I&-I&-I&I\end{pmatrix}
\end{equation}
on the ancillary qubits as described in equation (\ref{HadComb}), the qubits will be in the final state
\begin{equation}\begin{aligned}
    &H_1\otimes H_2 \hat{O}_P\ket{\phi_a}\ket{\psi_l}\\
    =&\frac{1}{2}\begin{pmatrix}I&I&I&I\\I&-I&I&-I\\I&I&-I&-I\\I&-I&-I&I\end{pmatrix}\frac{1}{B}\begin{pmatrix}\beta_0\begin{pmatrix}\alpha_{0l0}\\\alpha_{0l1}\end{pmatrix}\\\beta_1\begin{pmatrix}\alpha_{0l0}\\-\alpha_{0l1}\end{pmatrix}\\\beta_2\begin{pmatrix}\alpha_{0l1}\\\alpha_{0l0}\end{pmatrix}\\-i\beta_3\begin{pmatrix}\alpha_{0l1}\\-\alpha_{0l0}\end{pmatrix}\end{pmatrix}\\
    =&\frac{1}{2B}\begin{pmatrix}\left(\beta_0+\beta_1\right)\alpha_{0l0}+\left(\beta_2-i\beta_3\right)\alpha_{0l1}\\\left(\beta_2+i\beta_3\right)\alpha_{0l0}+\left(\beta_0-\beta_1\right)\alpha_{0l1}\\\left(\beta_0-\beta_1\right)\alpha_{0l0}+\left(\beta_2+i\beta_3\right)\alpha_{0l1}\\\left(\beta_2-i\beta_3\right)\alpha_{0l0}+\left(\beta_0+\beta_1\right)\alpha_{0l1}\\\left(\beta_0+\beta_1\right)\alpha_{0l0}-\left(\beta_2-i\beta_3\right)\alpha_{0l1}\\-\left(\beta_2+i\beta_3\right)\alpha_{0l0}+\left(\beta_0-\beta_1\right)\alpha_{0l1}\\\left(\beta_0-\beta_1\right)\alpha_{0l0}-\left(\beta_2+i\beta_3\right)\alpha_{0l1}\\-\left(\beta_2-i\beta_3\right)\alpha_{0l0}+\left(\beta_0+\beta_1\right)\alpha_{0l1}\end{pmatrix}\text{,}
\end{aligned}\end{equation}
where the coefficients of the states $\ket{000}$ and $\ket{001}$ returns the required results from equation (\ref{expect}), up to a normalization constant $\frac{1}{2B}$.

\subsection{Quantum Circuits}
\begin{figure}[htb]
    \centering
    \includegraphics[width=\textwidth]{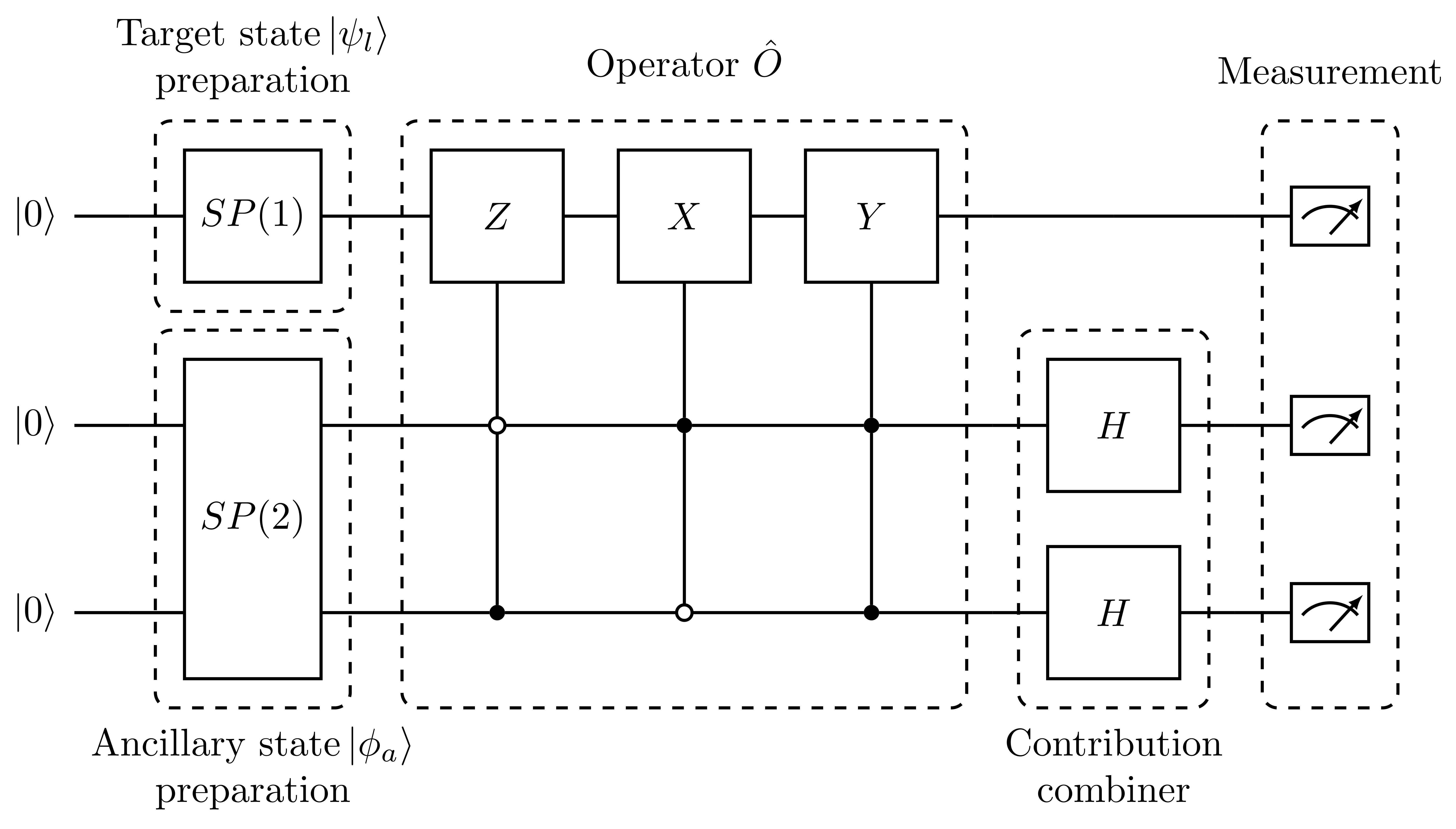}
    \caption{Quantum circuit performing one iteration of ITE in the $N=1$ case.}
    \label{QCN=1}
\end{figure}

\begin{figure}[H]
    \centering
    \includegraphics[width=\textwidth]{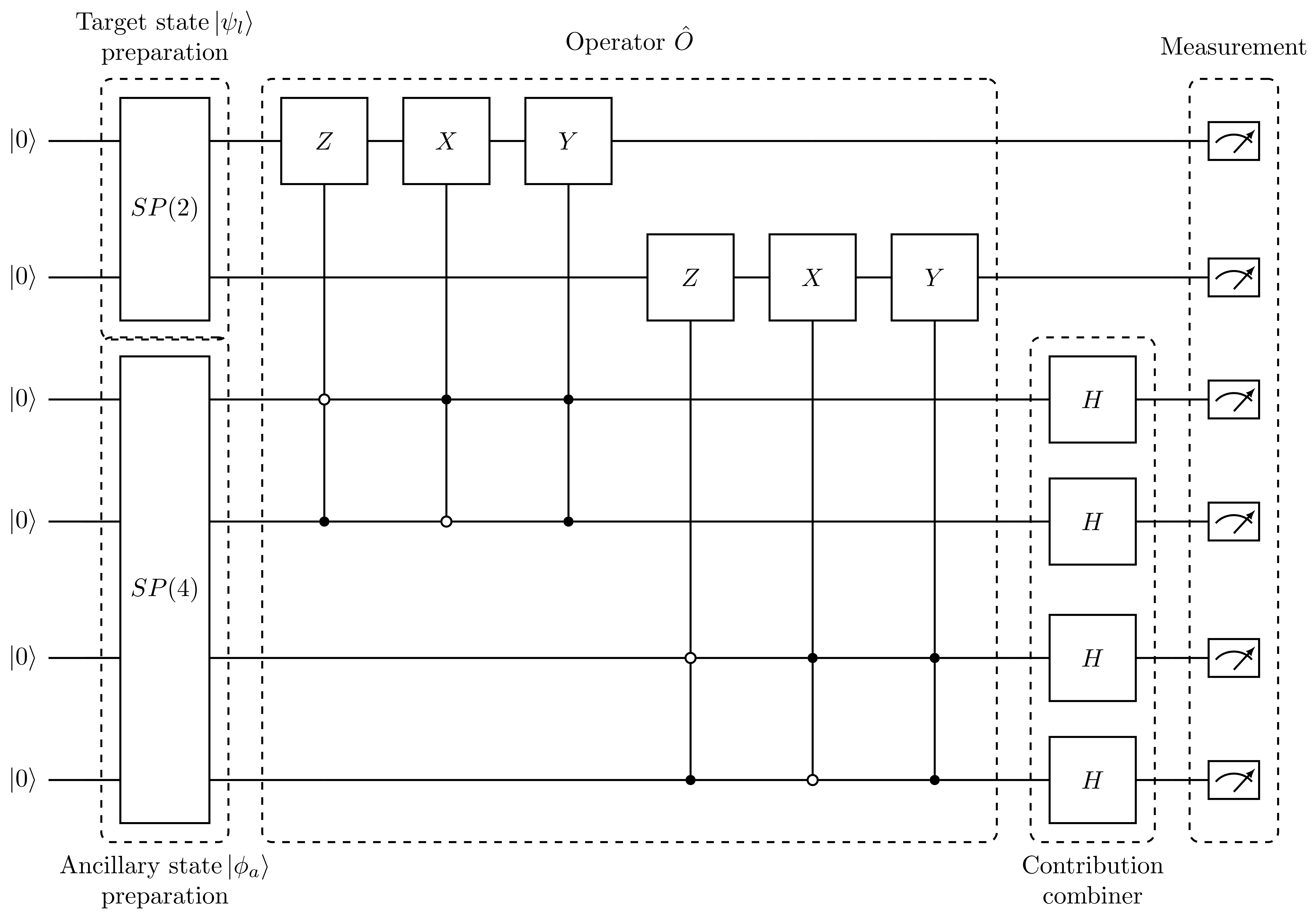}
    \caption{Quantum circuit performing one iteration of ITE in the $N=2$ case.}
    \label{QCN=2}
\end{figure}

For a state $\ket{\psi_l}$ expanded in the first $2^N$ basis states, the QITE algorithm can be performed by a $3N$ qubit circuit.  The $N=1$ and $N=2$ quantum circuits, which perform one iteration of imaginary time evolution, are shown in FIG. \ref{QCN=1} and \ref{QCN=2} respectively.

\onecolumngrid

\twocolumngrid

\begin{figure}[H]
    \centering
    \includegraphics[width=\textwidth]{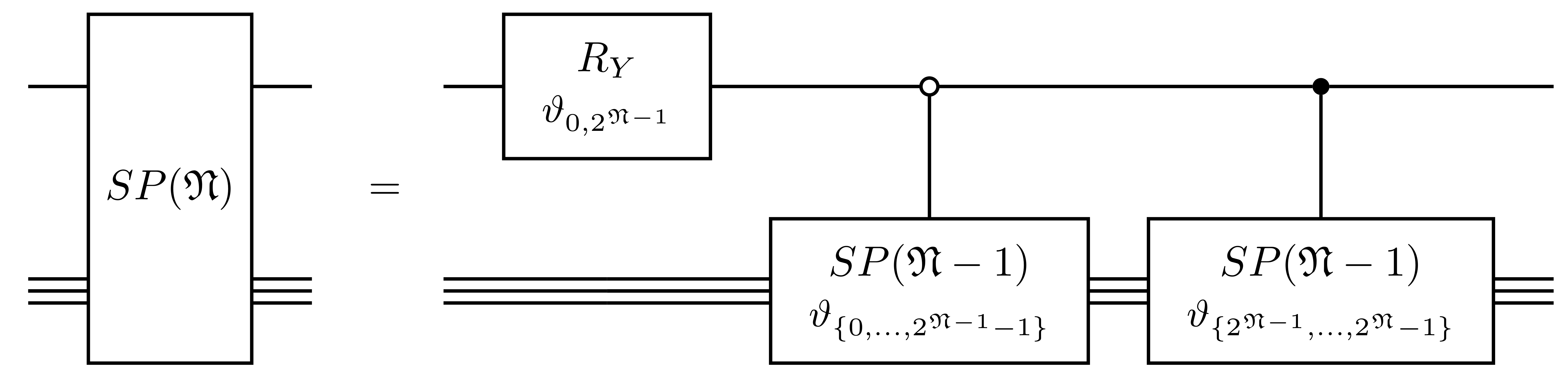}
    \caption{Quantum subcircuit $SP(\mathfrak{N})$ performing $\mathfrak{N}$-qubit real state preparation.}
    \label{SPN}
\end{figure}

\begin{figure}[H]
    \centering
    \includegraphics[width=\columnwidth]{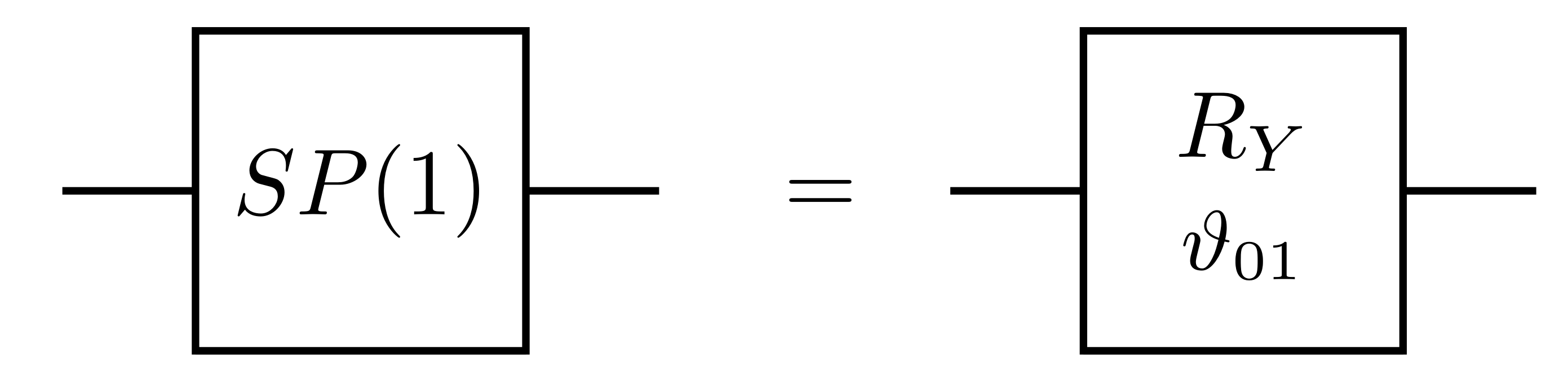}
    \caption{Quantum subcircuit $SP(1)$ performing one-qubit real state preparation.}
    \label{SP1}
\end{figure}

Here the open circles and filled circles indicate control activation if the control qubit is in the state $|0\rangle$ and $|1\rangle$ respectively. FIG. \ref{SPN} shows the subcircuit $SP(\mathfrak{N})$ for a $\mathfrak{N}$-qubit real state preparation, where the one-qubit ($\mathfrak{N}=1$) state preparation subcircuit $SP(1)$ and $SP(2)$ is given by FIG. \ref{SP1}.

\onecolumngrid

\twocolumngrid

In a $\mathfrak{N}$-qubit real state there are $2^\mathfrak{N}$ coefficients ($2^\mathfrak{N}-1$ independent ones), denoted $\{c_i\}$, $i=0,1,\dots,2^\mathfrak{N}-1$. For the target state $\ket{\psi_l}$ preparation subcircuits $SP(N)$, $c_{n'}=\alpha_{0ln'}$ (as in equation \ref{decomstate}). For the ancillary state $\ket{\phi_a}$ preparation subcircuits $SP(2N)$, $c_{i}=\beta_{i}$ (as in equation \ref{decomH}). The angles of rotation, $\vartheta_{ij}$, for the state preparation, are given by
\begin{equation}
    \tan{\frac{\vartheta_{ij}}{2}}=\sqrt{\frac{\sum_{i'=j}^{2j-i-1}c_{i'}^2}{\sum_{i'=i}^{j-1}c_{i'}^2}}\text{.}
\end{equation}

\begin{figure}[H]
    \centering
    \includegraphics[width=\columnwidth]{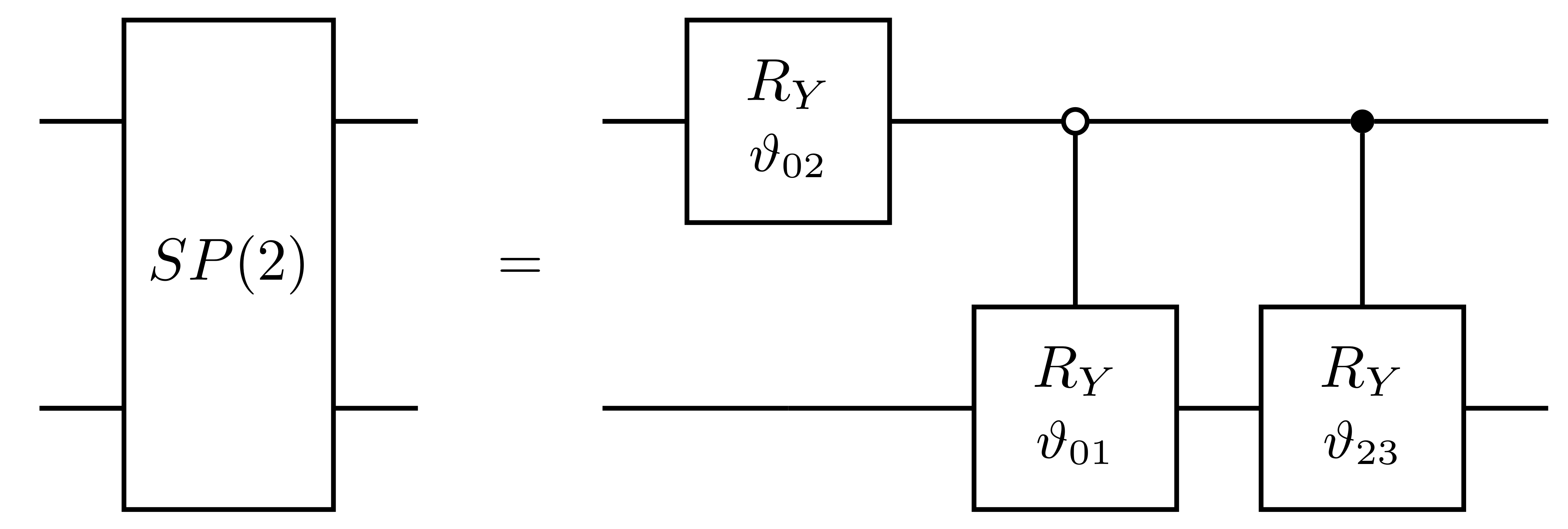}
    \caption{Quantum subcircuit $SP(2)$ performing two-qubit real state preparation.}
    \label{SP2}
\end{figure}

For $\mathfrak{N}=2$, $SP(2)$ is shown in FIG. \ref{SP2}, with the angles given by
\begin{equation}\begin{aligned}
    \tan{\frac{\vartheta_{02}}{2}}&=\sqrt{\frac{c_2^2+c_3^2}{c_0^2+c_1^2}}\\
    \tan{\frac{\vartheta_{01}}{2}}&=\frac{c_1}{c_0}\\
    \tan{\frac{\vartheta_{23}}{2}}&=\frac{c_3}{c_2}\text{.}
\end{aligned}\end{equation}

In general the statevector is not accessible through measurements as only the relative amplitudes can be obtained through the probabilties. In our algorithm we exploit the fact that all of our components are real and the relative phases can either be $0$ or $\pi$. After direct measurements for deducing the relative amplitudes of the states, $N$ more circuits are composed by appending a Hadamard gate to the $N\textsuperscript{th}$ target qubit. This mixes the amplitudes. An increase in probablity indicates a relative phase of $0$ between the two states, and a decrease indicates a relative phase of $\pi$.

\section{Results}
\subsection{$N=1$ case}
We applied the approach above on a 3 qubit simulator ($N=1$, 2 expansion states) to calculate the ground state energy of \ce{^{4}He} \cite{PhysRev.123.420,JChemPhys.148.094101}, where all the nucleons are in the $s$ state ($l=0$) with spin and isospin degeneracy $=4$, using the parameter values $t_0=-1090.0\,\si{\MeV\femto\m\cubed}$ and $t_3=17288.0\,\si{\MeV\femto\m\tothe{6}}$ \cite{PhysLettB.61.227}.

\begin{figure}[tbh]
    \subfloat[\label{C1wfn}]{\includegraphics[width=\columnwidth]{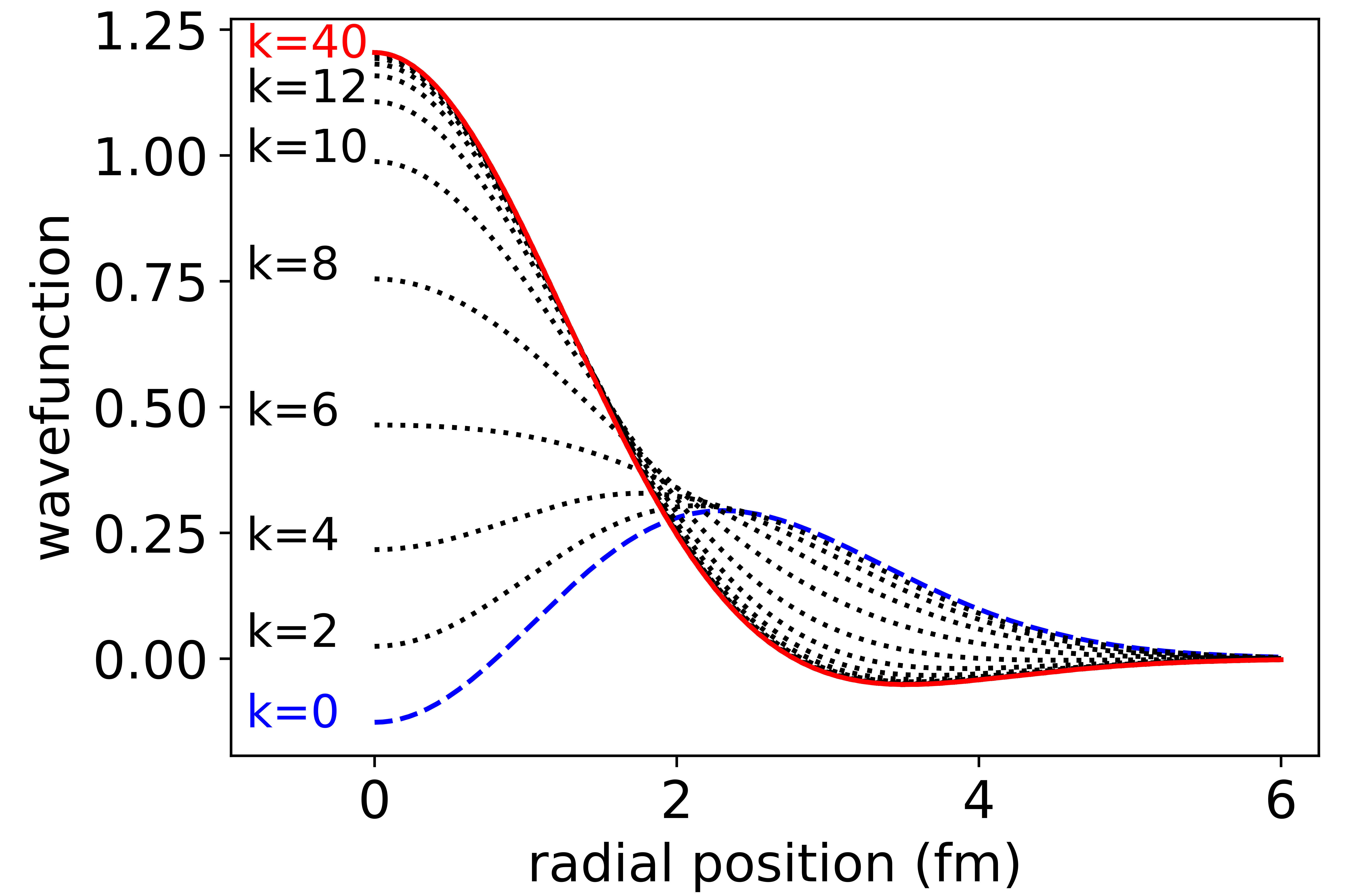}}
    
    \subfloat[\label{C1pot}]{\includegraphics[width=\columnwidth]{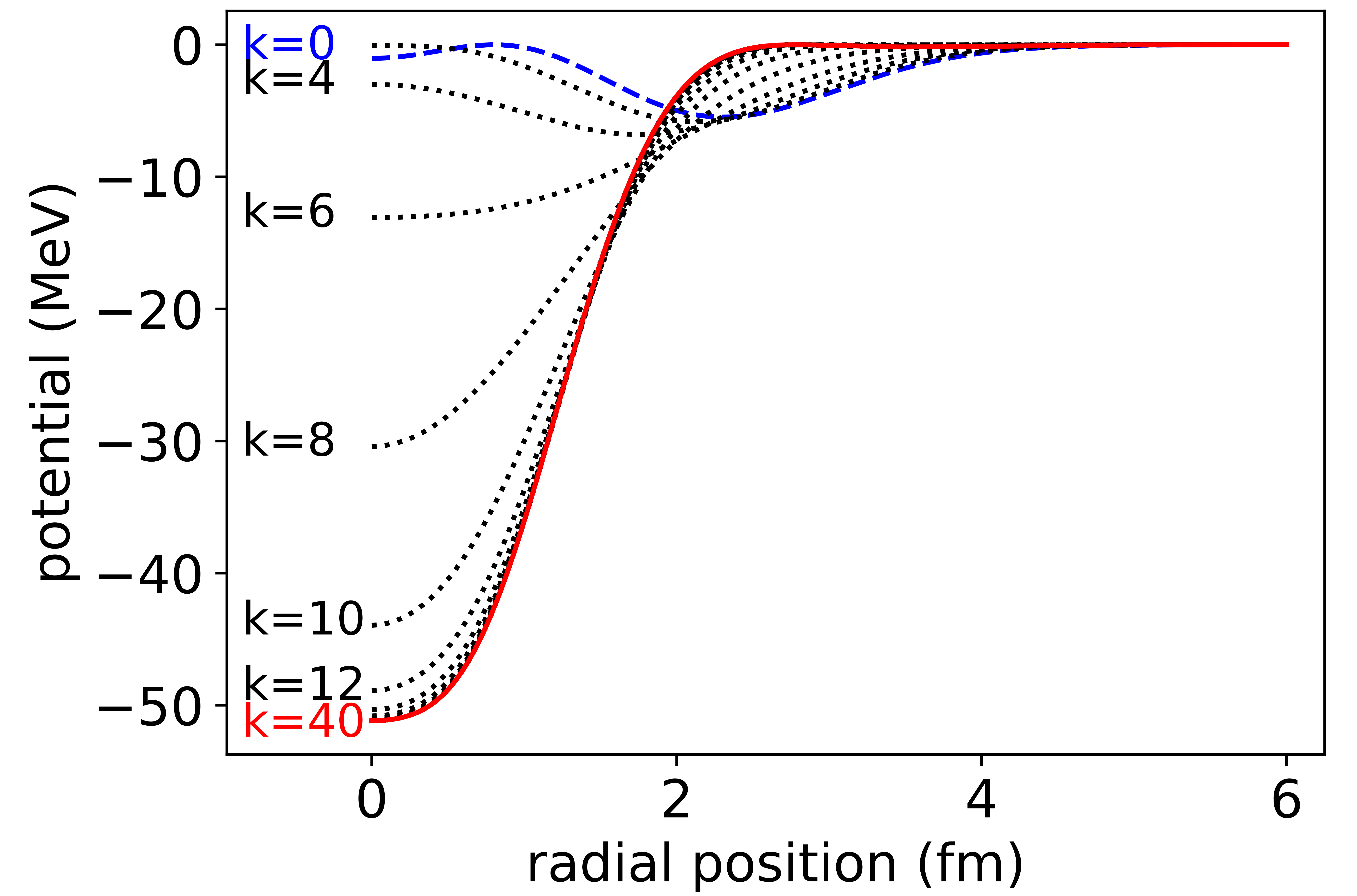}}
    \caption{Classical imaginary time evolution for $N=1$.  (a) Wave function as a function of iteration.  (b) Potential as a function of iteration.}
    \label{C1_0.005_1.53}
\end{figure}

We choose a time step of $\frac{\Delta\tau}{\hbar}=0.005\,\si{\MeV\tothe{-1}}$ and use an optimized oscillator length of $\frac{1}{b}=1.5284\,\si{\femto\m}$.  Then we perform an ITE, using a classical algorithm, from an initial trial state with equal amplitudes of the first two oscillator states,
\begin{equation}
    u_{00}^{(0)}=\frac{1}{\sqrt{2}}\mathcal{R}^{\frac{1}{b}, \frac{3}{2}}_{0}+\frac{1}{\sqrt{2}}\mathcal{R}^{\frac{1}{b}, \frac{3}{2}}_{1}\text{.}
\end{equation}
The wave function starts achieving self-consistency (up to 3 significant figures) after 20 iterations. Further evolution (40 iterations) of the state gives a \ce{^{4}He} ground state energy of $-29.1567\,\si{\MeV}$. The precision of the energy here indicates the self-consistency of the value over the last 3 iterations. In this case further iteration only changes digits from the fifth decimal place. FIG. \ref{C1_0.005_1.53} shows the evolution of the wave function and the potential as a function of iteration.  Even iteration numbers only are shown for clarity. 

\begin{figure}[tbh]
    \subfloat[\label{Qm1wfn}]{\includegraphics[width=\columnwidth]{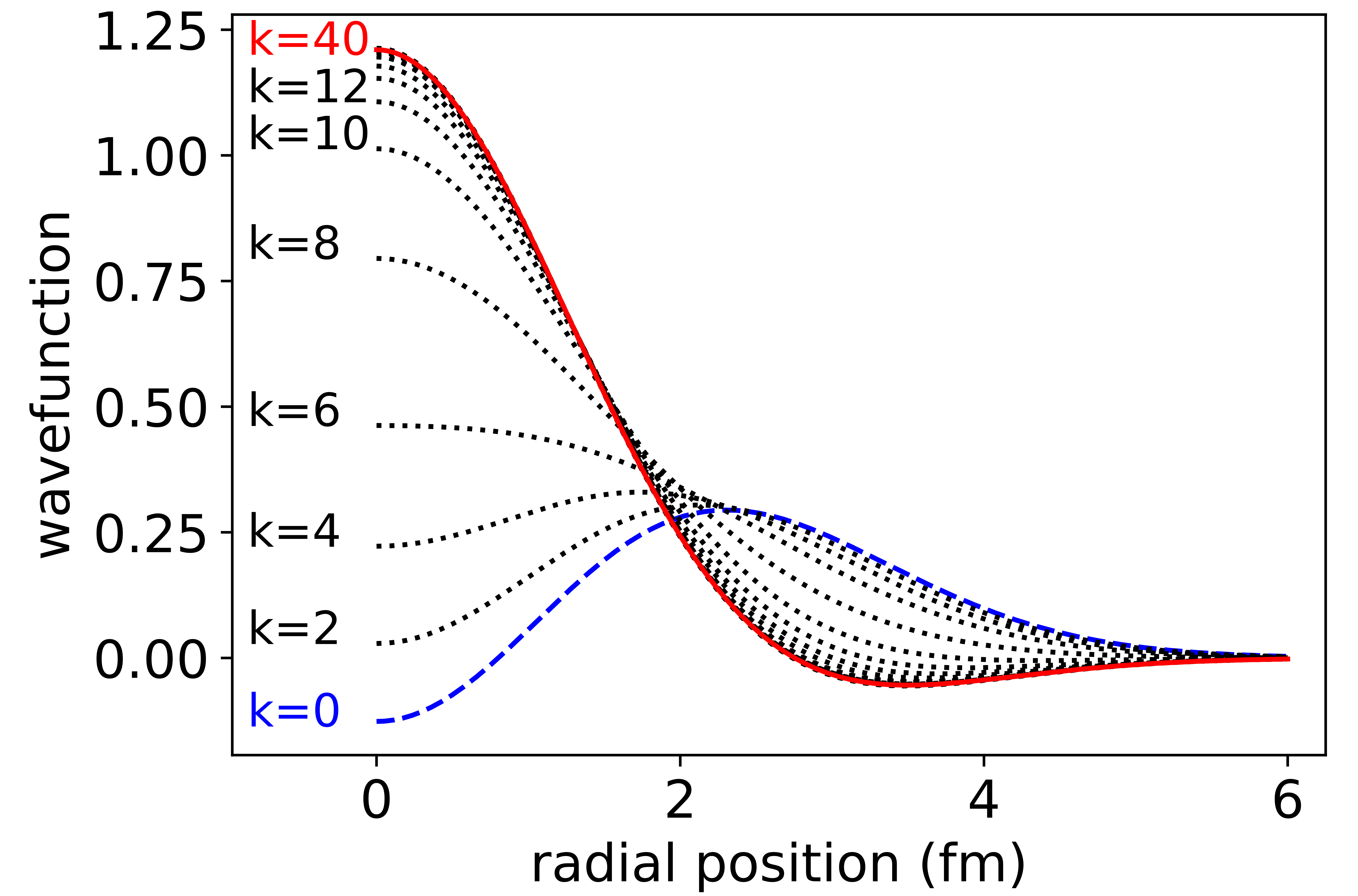}}
    
    \subfloat[\label{Qm1pot}]{\includegraphics[width=\columnwidth]{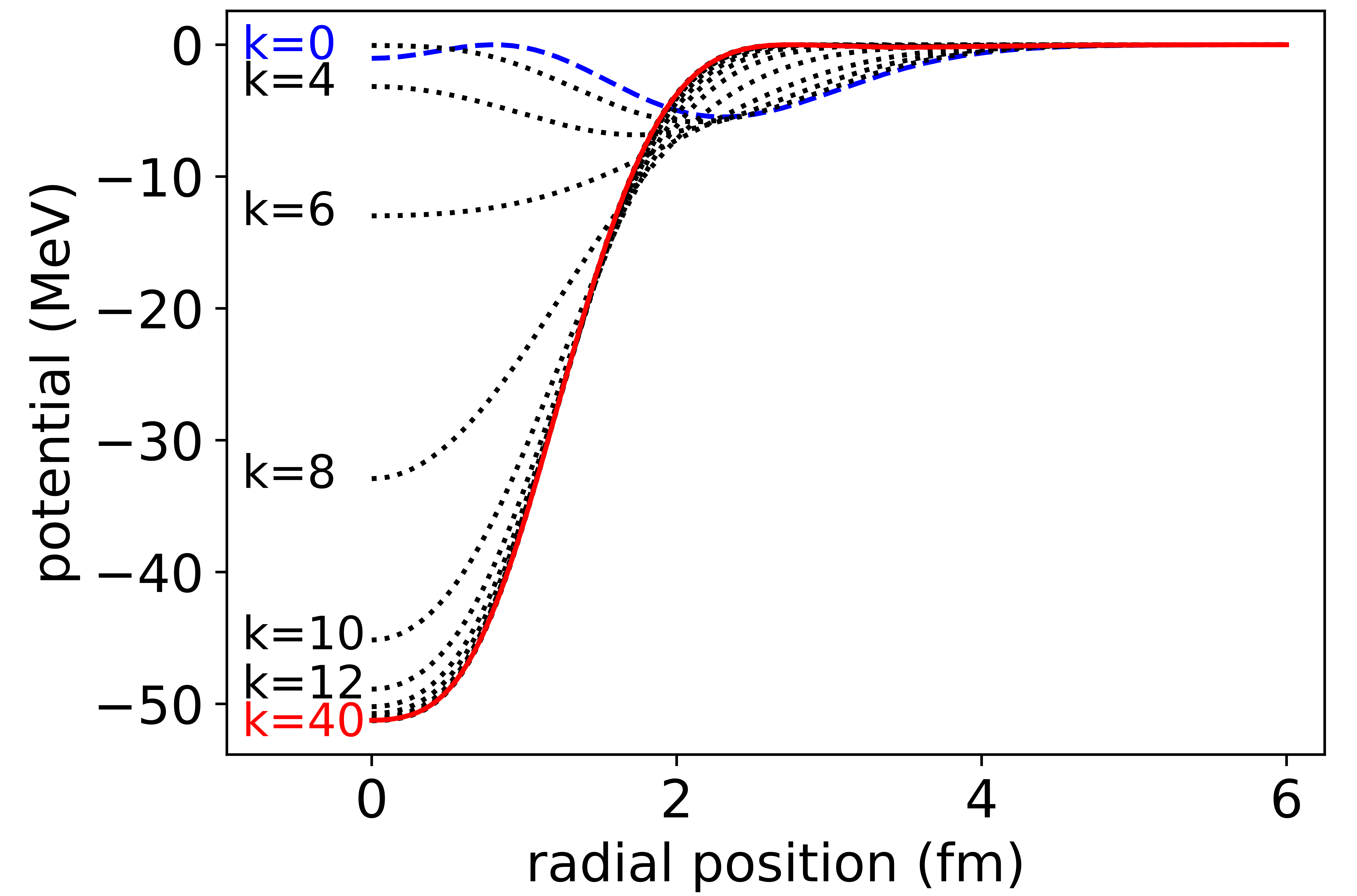}}
    \caption{Quantum imaginary time evolution for $N=1$ carried out using a quantum simulator. (a) Wave function as a function of iteration.  (b) Potential as a function of iteration.}
    \label{Qm1_0.005_1.53_10000}
\end{figure}

\noindent The quantum imaginary time evolution procedure was coded in Qiskit \cite{MTreinish.2023} and simulated on a classical computer using the QASM backend with $10000$ shots per measurement. The same starting wave function, oscillator length, and number of time steps were used as in the classical state for comparison. FIG. \ref{Qm1_0.005_1.53_10000} shows the evolution of the wave function and potential as a function of iteration for this implementation of the quantum imaginary time evolution algorithm.  The final ground state energy obtained was $-29.14\,\si{\MeV}$. Self-consistency is only achieved up to 2 decimal places after 40 iterations.

\begin{figure}[tbh]
    \centering
    \includegraphics[width=\columnwidth]{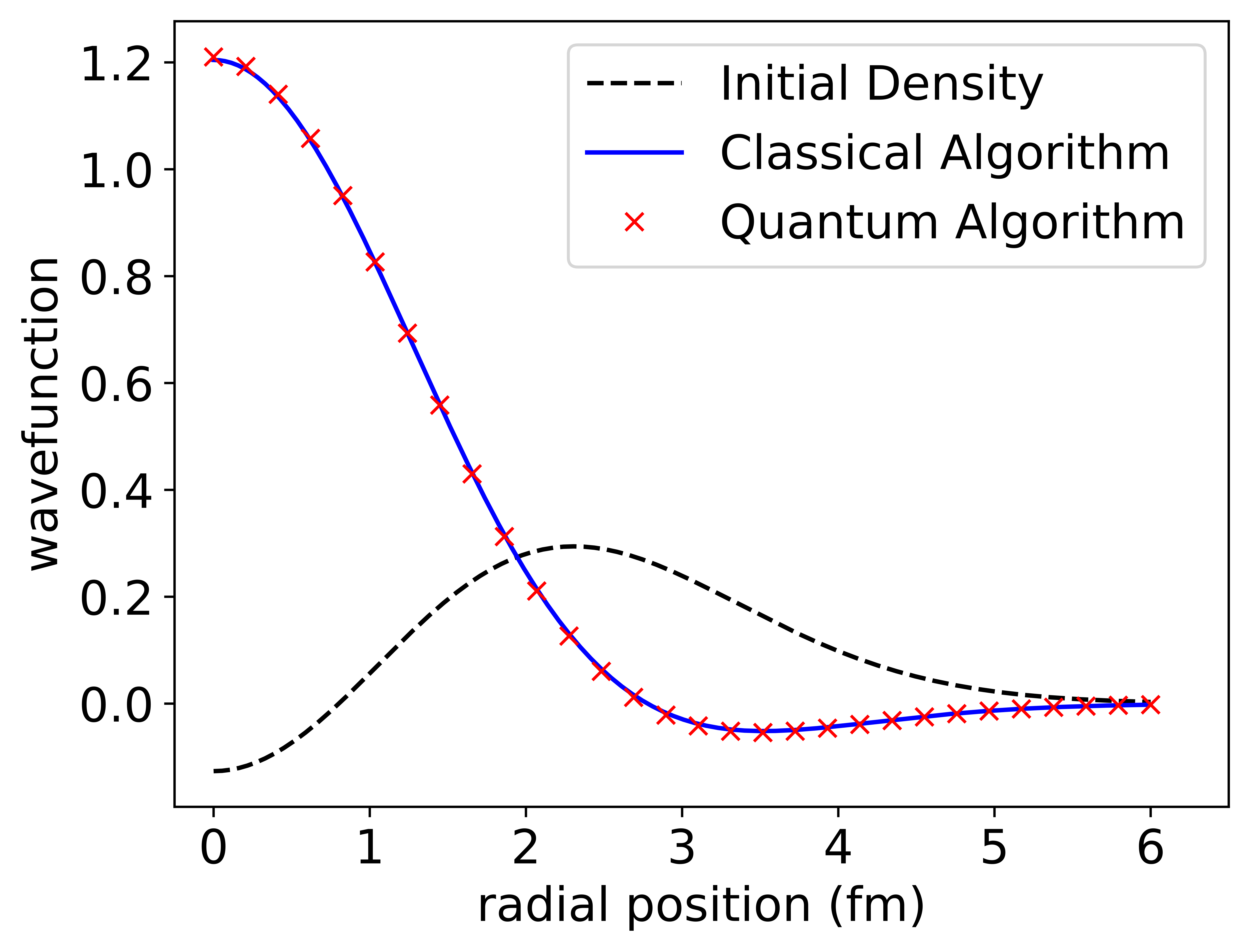}
    \caption{Imaginary time evolution for $N=1$ case after 40 iterations.}
    \label{CQ1_40}
\end{figure}

\noindent FIG. \ref{CQ1_40} shows the final ground state wave functions after 40 iterations from classical and quantum ITE.  The per-iteration results and the final energy are thus in close agreement with the classical implementation, as expected.

\subsection{$N=2$ case}
We repeated the above procedures on a 6 qubit simulator ($N=2$, 4 expansion states), without changing any parameters apart from the oscillator lengths and the initial trial state. The initial state is chosen to have equal amplitudes of the first four oscillator states,
\begin{equation}
    u_{00}^{(0)}=\frac{1}{2}\mathcal{R}^{\frac{1}{b}, \frac{3}{2}}_{0}+\frac{1}{2}\mathcal{R}^{\frac{1}{b}, \frac{3}{2}}_{1}+\frac{1}{2}\mathcal{R}^{\frac{1}{b}, \frac{3}{2}}_{2}+\frac{1}{2}\mathcal{R}^{\frac{1}{b}, \frac{3}{2}}_{3}\text{,}
\end{equation}
while the oscillator lengths are optimized to get the lowest ground state energies. The values we use is $\frac{1}{b}=2.0635\,\si{\femto\m}$.

\noindent Compared to the $N=1$ case, more iterations are needed for the wave function to achieve self-consistency. Under the classical algorithm, the wave function achieves self-consistency (up to 3 significant figures) after 35 iterations. We evolved the state for an addition of 25 iterations to obtain a \ce{^{4}He} ground state energy of $-34.8215\,\si{\MeV}$. 

\begin{figure}[H]
    \subfloat[\label{Qm2wfn}]{\includegraphics[width=\columnwidth]{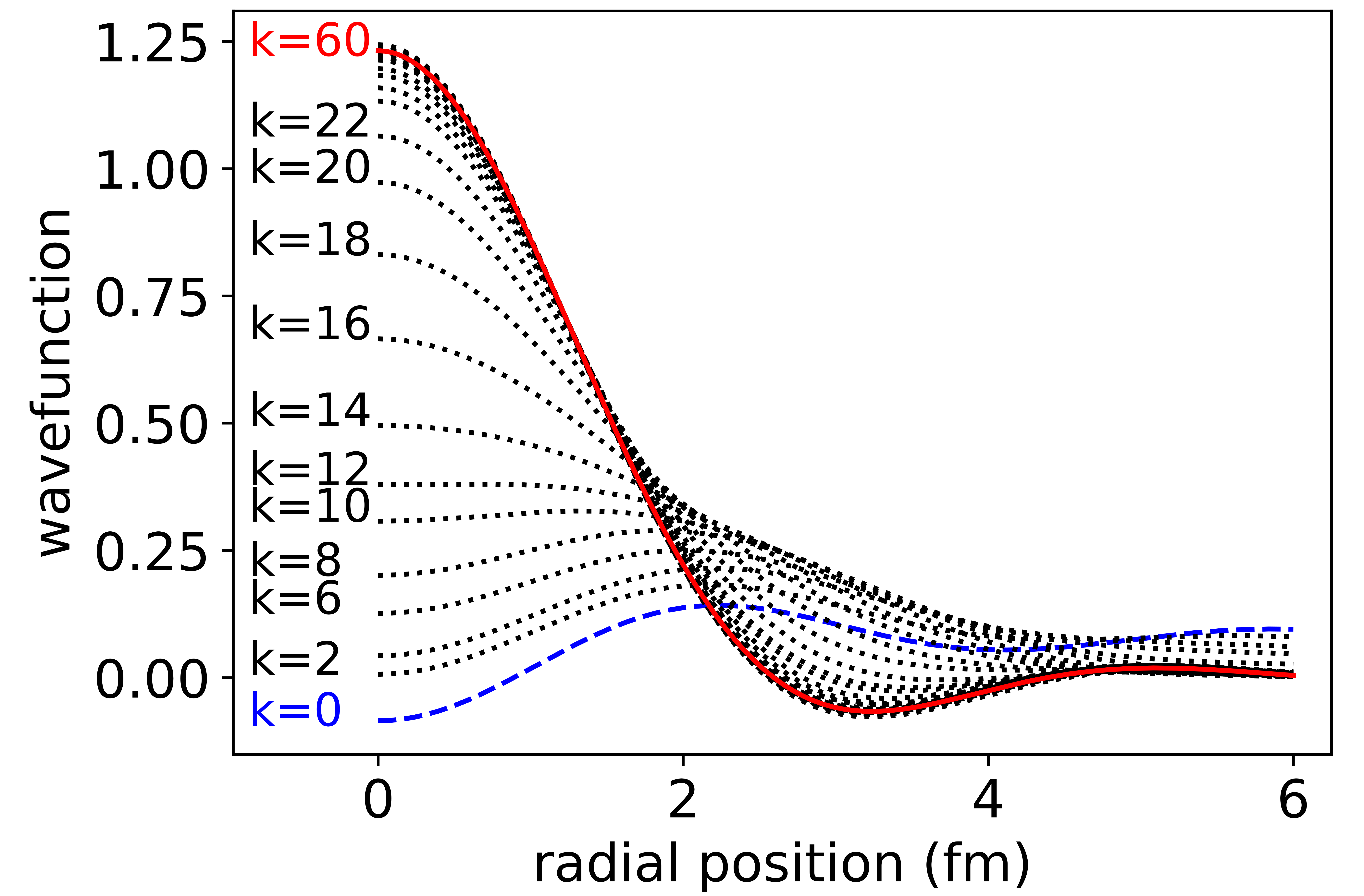}}
    
    \subfloat[\label{Qm2pot}]{\includegraphics[width=\columnwidth]{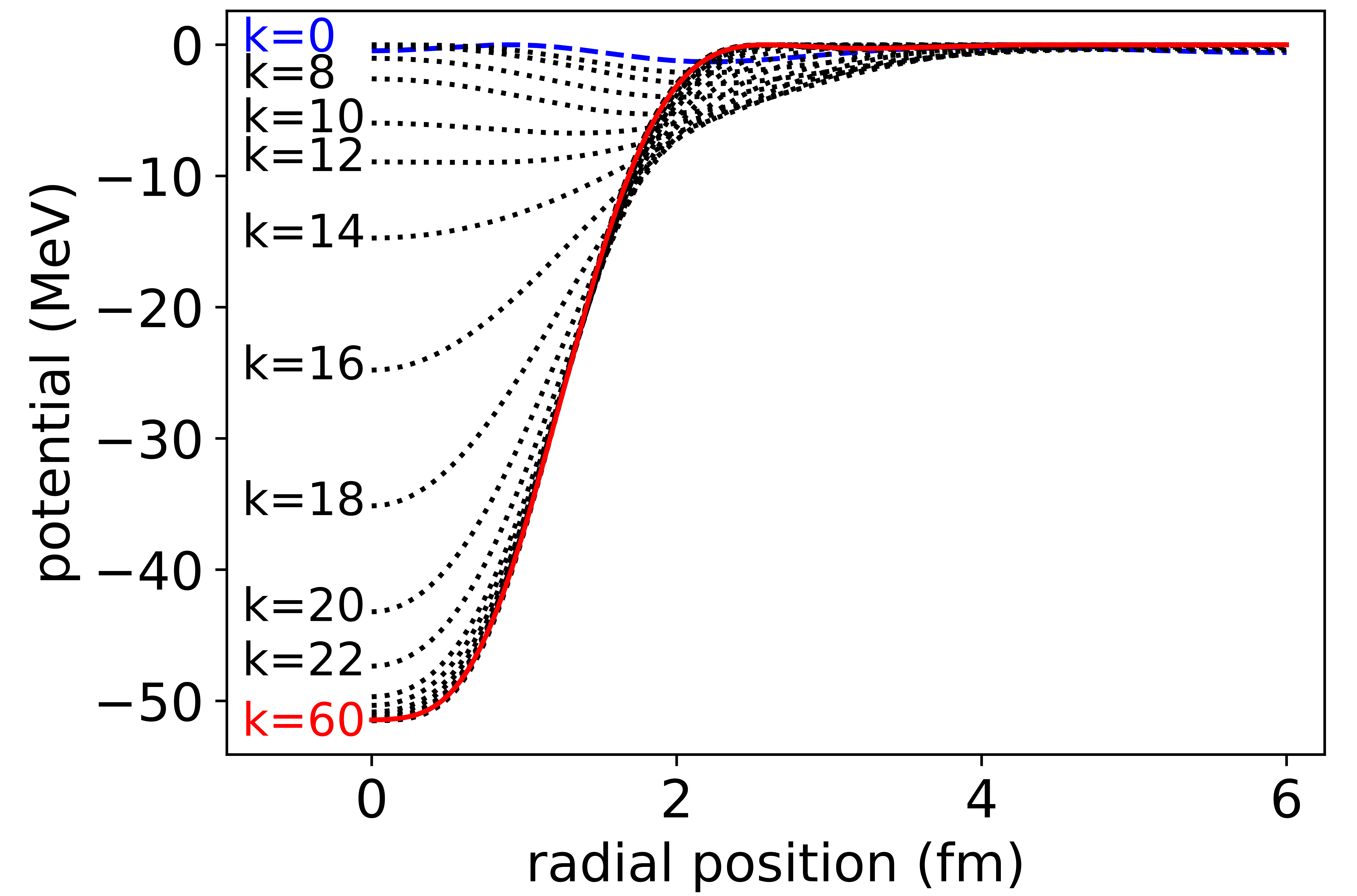}}
    \caption{Quantum imaginary time evolution for $N=2$ carried out using a quantum simulator. (a) Wave function as a function of iteration. (b) Potential as a function of iteration.}
    \label{Qm2_0.005_2.06_10000}
\end{figure}

\noindent The wave function evolved under the QITE implementation achieved self-consistency (up to 3 significant figures) after 50 iterations. The final (60 iterations) \ce{^{4}He} ground state energy obtained was $-34.8\,\si{\MeV}$. The evolution is showed in FIG. \ref{Qm2_0.005_2.06_10000}.

\begin{figure}[h]
    \centering
    \includegraphics[width=\columnwidth]{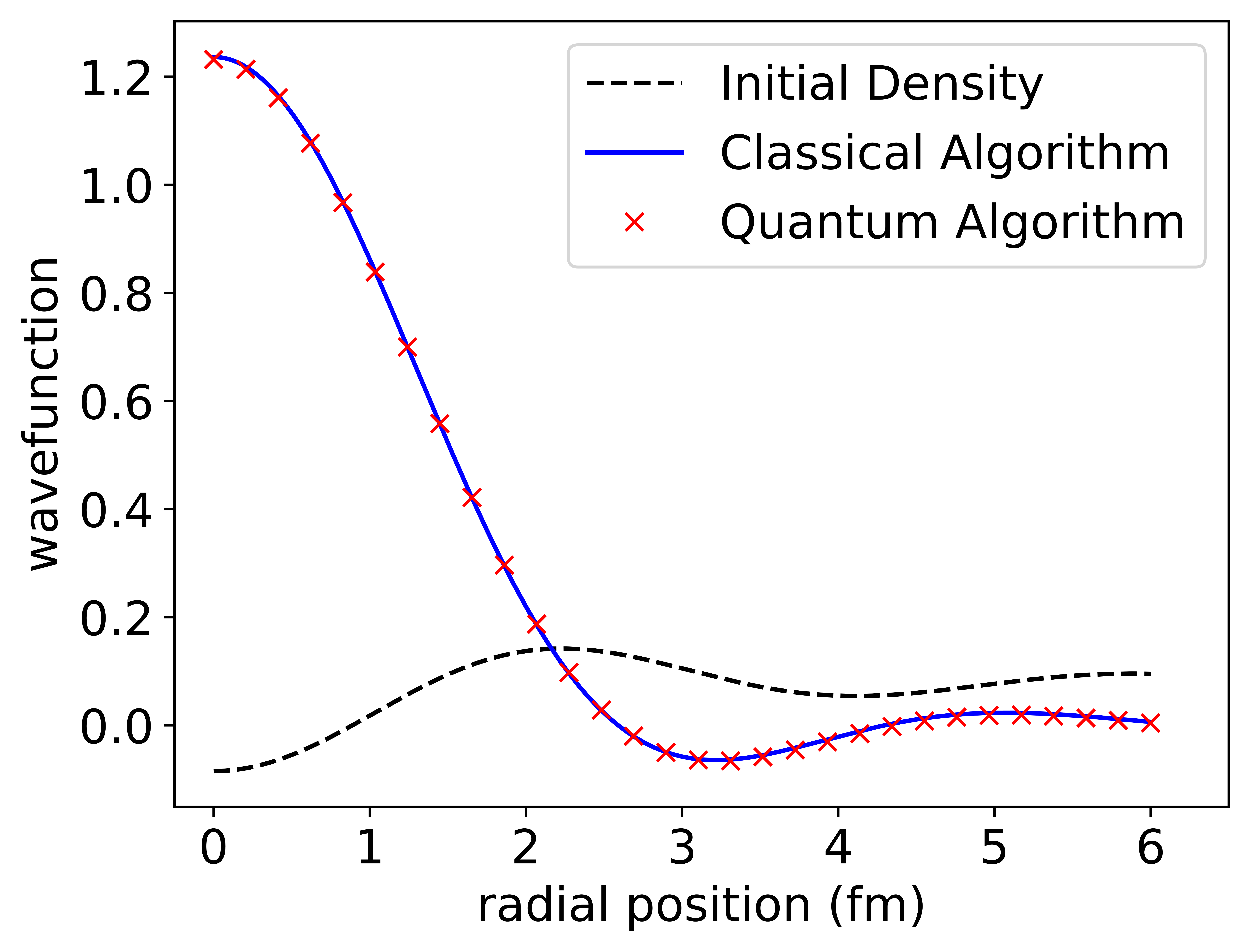}
    \caption{Imaginary time evolution for $N=2$ case after 60 iterations.}
    \label{CQ2_60}
\end{figure}

\noindent FIG. \ref{CQ2_60} shows the final ground state wave functions after 60 iterations from classical and quantum ITE in the $N=2$ case. Compared to the $N=1$ case, the precision of the ground state energy is lower. Improvement of this can be done by using more number of shots per measurement and evolving the states for more iterations.

\subsection{Pauli Coefficients}
\begin{figure}[bh]
    \subfloat[\label{Qm1Paulifull}]{\includegraphics[width=\columnwidth]{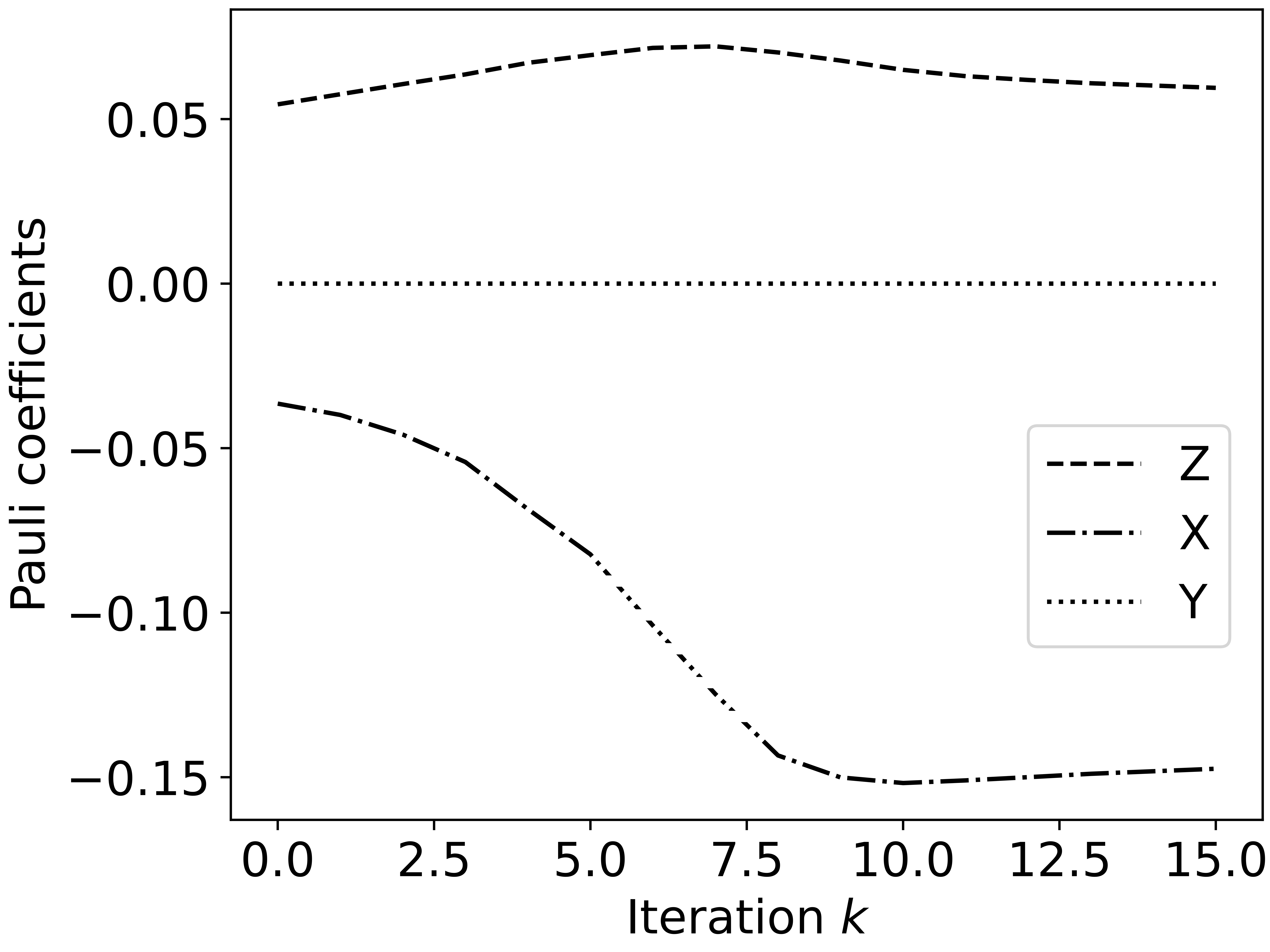}}
  
    \subfloat[\label{Qm1Paulizoom}]{\includegraphics[width=\columnwidth]{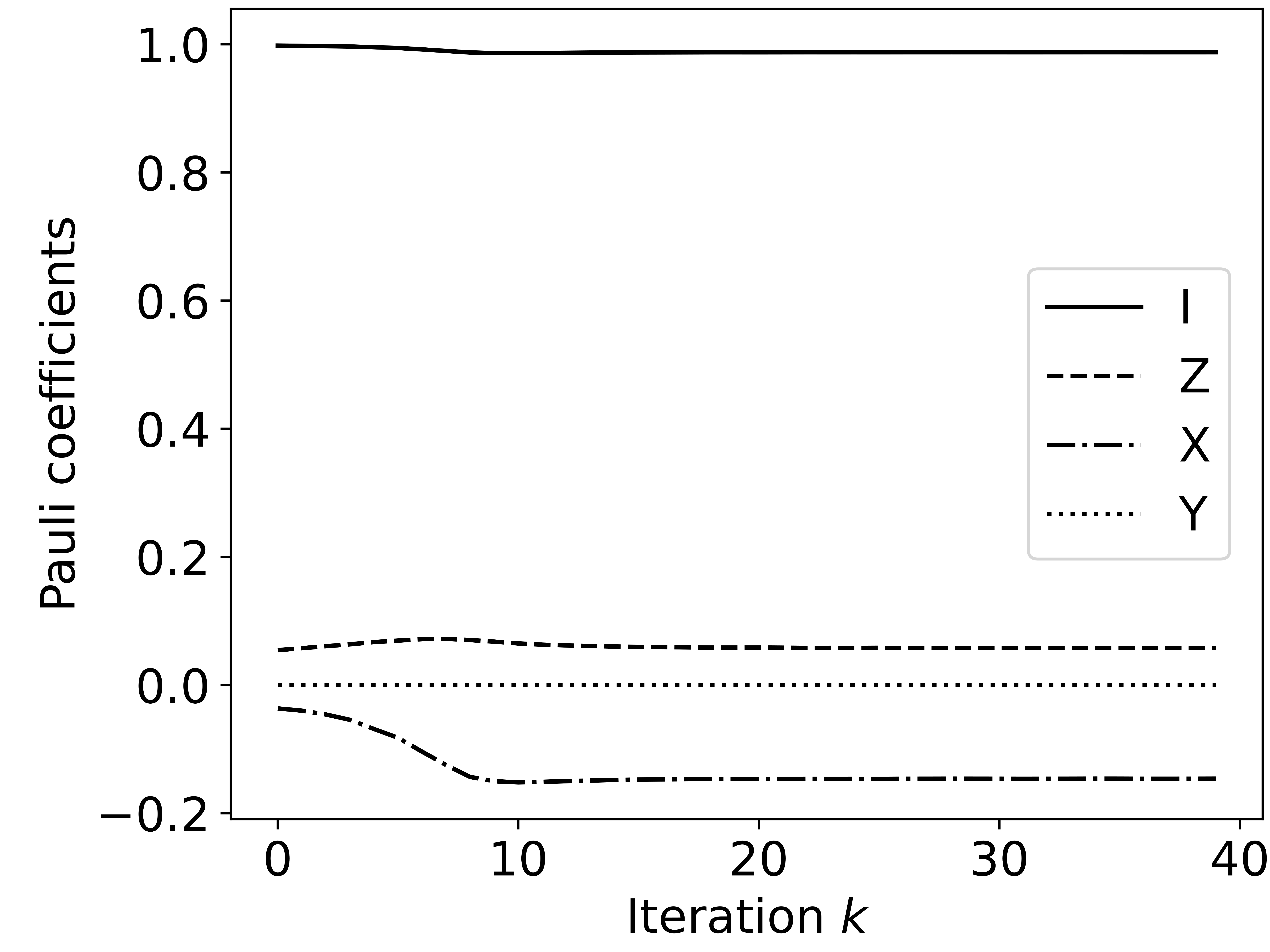}}
    \caption{Per-iteration development of the $\beta$ coefficients for the terms in the Pauli expansion of the unitary operator in the quantum imaginary time evolution algorithm for $N=1$ case. (a) Full view. (b) Zoomed in view.}
  \label{Qm1_0.005_1.53_10000_Pauli}
\end{figure}

\begin{figure}[H]
    \centering
    \includegraphics[width=\columnwidth]{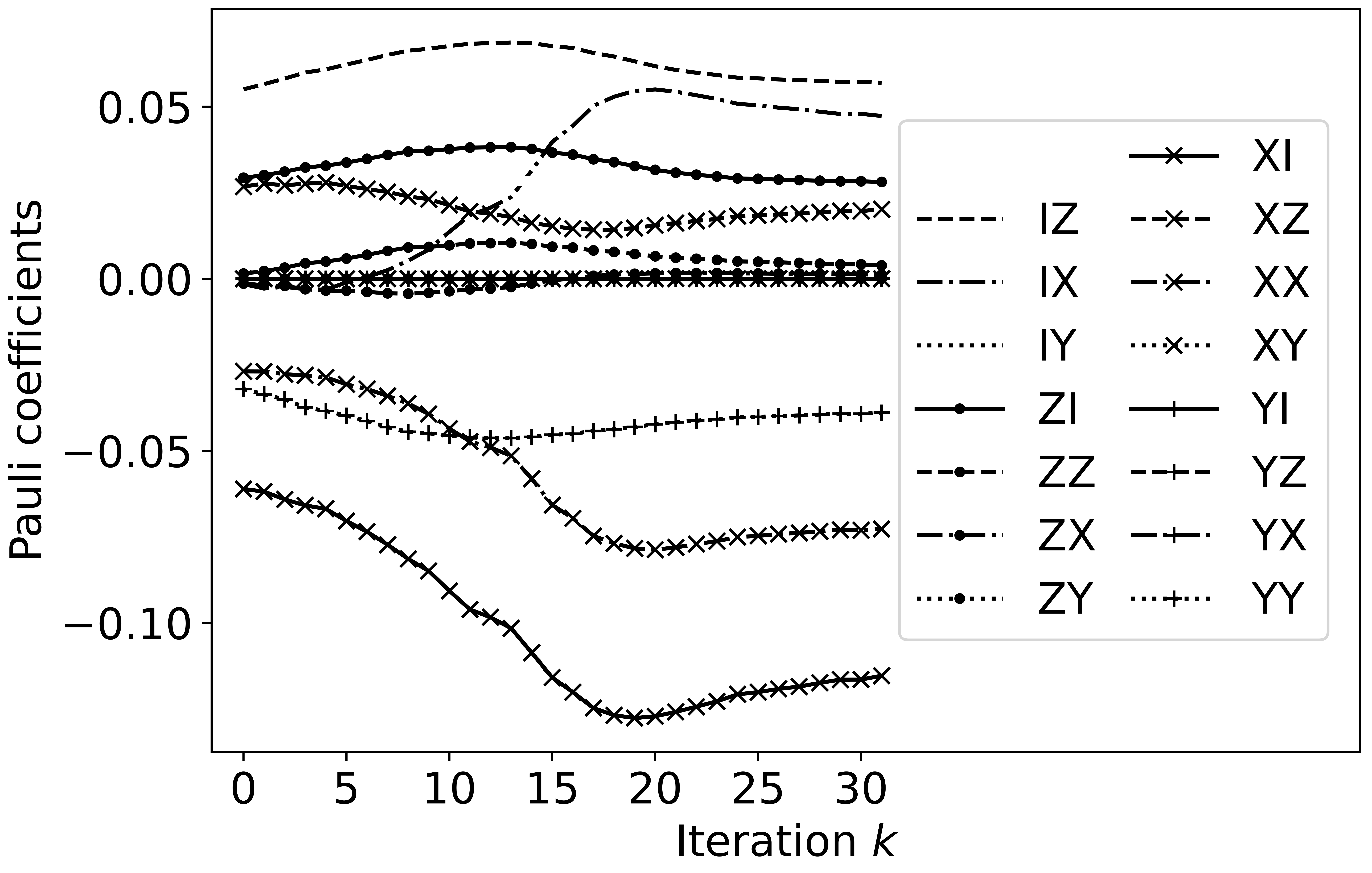}
    \caption{Per-iteration development of the $\beta$ coefficients for the terms in the Pauli expansion of the unitary operator in the quantum imaginary time evolution algorithm for $N=2$ case.}
  \label{Qm2_0.005_2.06_10000_Pauli}
\end{figure}

To gain some understanding of the operation of the quantum imaginary time algorithm, we examine the coefficients, $\beta_i$ of the Pauli operators in the expansion of the unitary operator $\hat{U}$ (\ref{decomH}).  The results are shown in FIG. \ref{Qm1_0.005_1.53_10000_Pauli} and \ref{Qm2_0.005_2.06_10000_Pauli}.

\noindent The $N=2$ case shows a similar behavior as the $N=1$ case, where the identity operator has a unperturbed contribution near 1.0 while the other operators have much lower contributions. The identity operator $II$ is not shown in the $N=2$ case (fig. \ref{Qm2_0.005_2.06_10000_Pauli}) for clarity.

\noindent We see the changing over iteration number of the structure of the imaginary time operator as the effective density-dependent Hamiltonian changes, until the operator converges to a fixed form. The relative importance of different terms may allow for a future optimised implementation which eliminates the smallest contributions, but we have not attempted that here.

\subsection{Quantum Scaling}
The present work uses a small basis expansion as a proof of concept, where the quantum advantage is not apparent. In this section we compare the computational cost in two different aspects, the number of bits/qubits used (circuit width) and the number of operations/gates used.

\begin{figure}[bh]
    \centering
    \includegraphics[width=\textwidth]{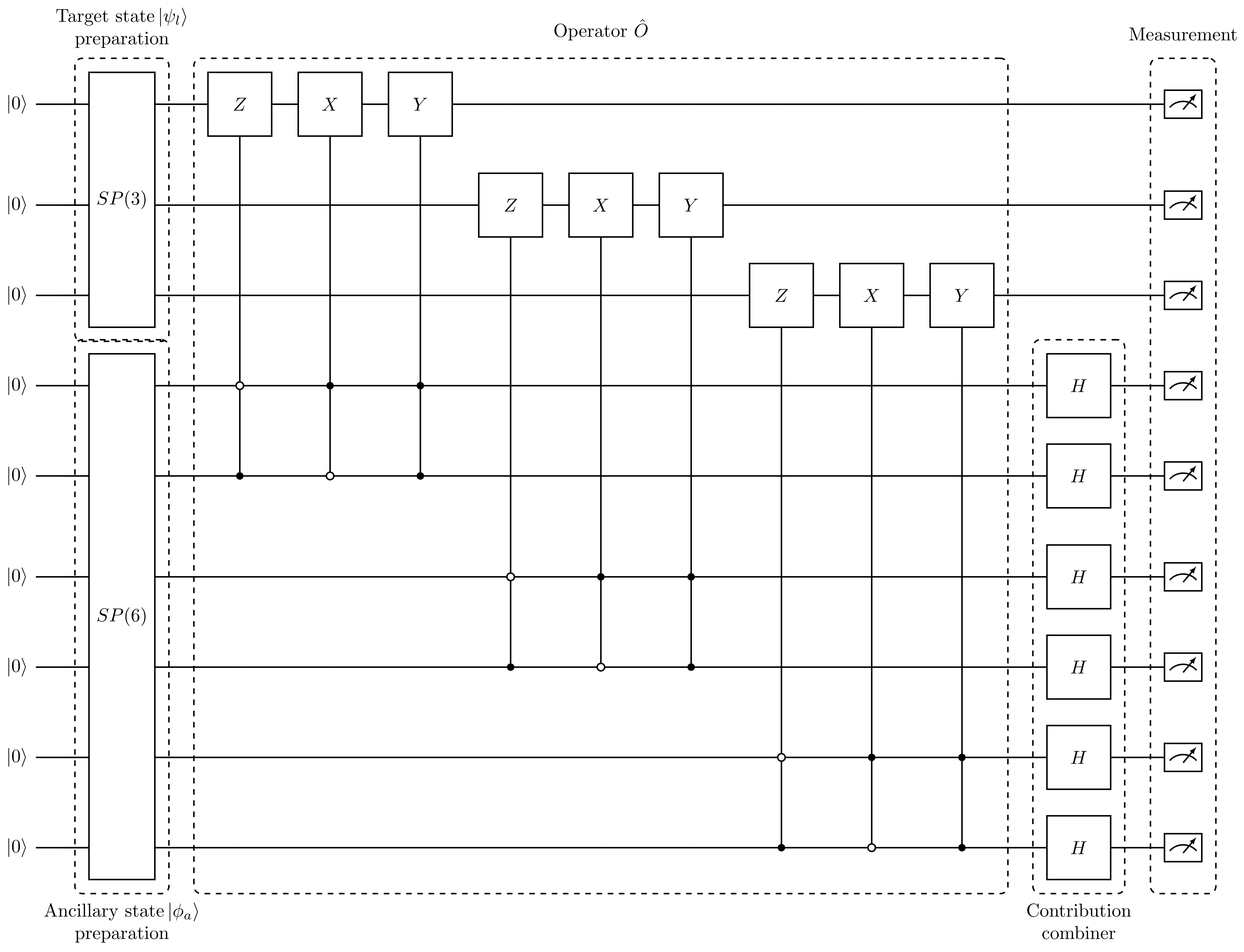}
    \caption{Quantum circuit for one time step in the $N=3$ case.}
    \label{QCN=3}
\end{figure}

\begin{figure}[H]
    \centering
    \includegraphics[width=\textwidth]{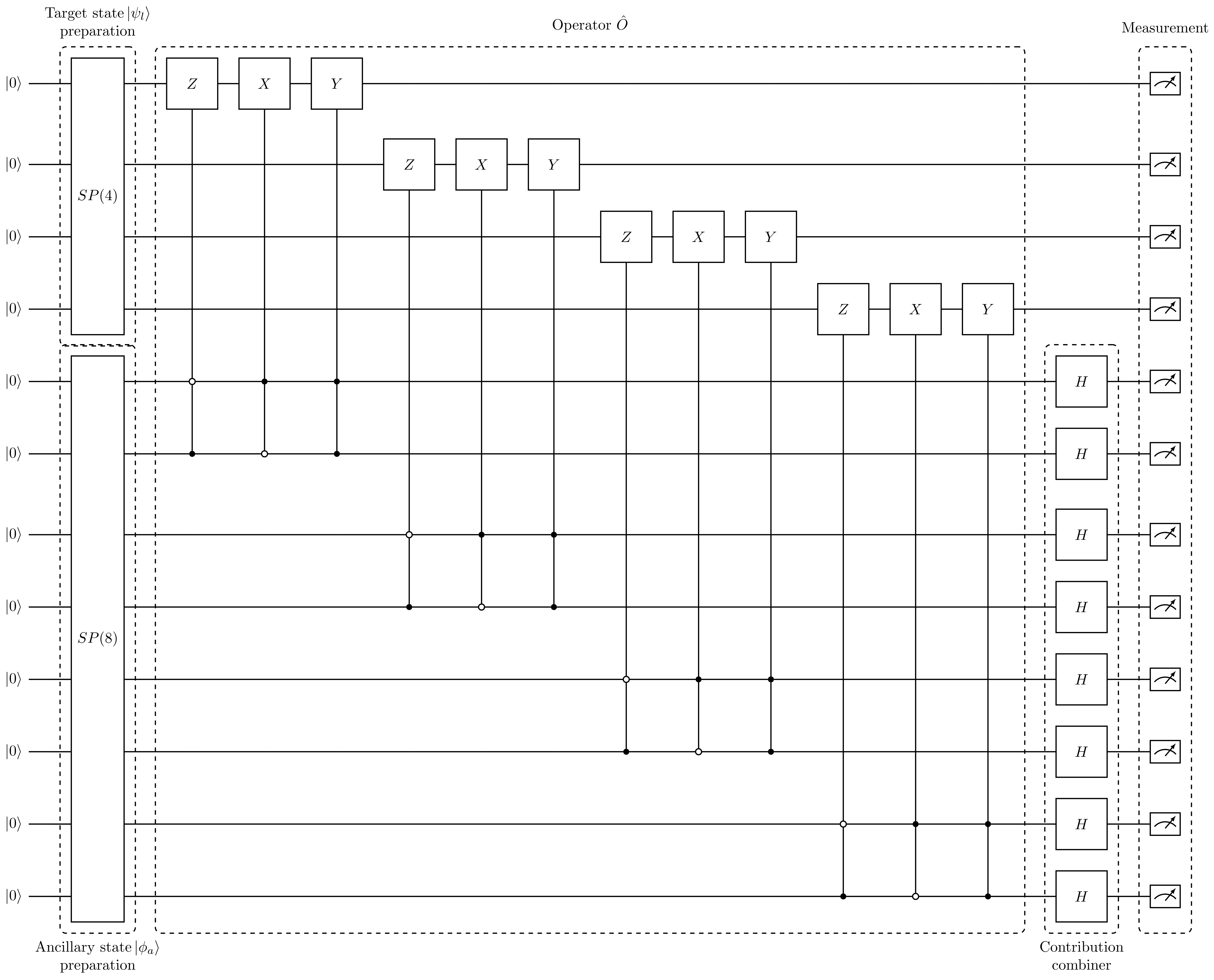}
    \caption{Quantum circuit for one time step in the $N=4$ case.}
    \label{QCN=4}
\end{figure}

With $2^N$ expansion states, the classical algorithm requires $2^N$ bits to set up, whereas its quantum counterpart uses only $3N$ qubits for the presented implementation of QITE. This improvement is exponential and allows one to carry out calculations with a much larger basis expansion size in the eventual availability of larger quantum computers. In our case, we have used a technique in which general non-unitary operators can be used. In theory only one ancillary qubit is needed to implement the circuit. In our algorithm the extra ancillary qubits are traded for a reduced circuit depth. The linear growth in circuit depth is demonstrated in FIG. \ref{QCN=3} and \ref{QCN=4}, which contain the quantum circuits for larger basis expansions. Specialisation to the QITE operator can reduce the ancillary qubit cost further.

\onecolumngrid

\twocolumngrid

In terms of the size of the algorithm, the classical ITE algorithm requires $O\left(2^{2N}\right)$ operations per time step. One common measure of the complexity of a quantum ciruit is the number of multi-qubit gates in it. As all quantum gates can be decomposed into 1-qubit gates and the 2-qubit CNOT gates, this is equivalent to counting the number of CNOT gates.

Our quantum circuit (FIG. \ref{QCN=1}, \ref{QCN=2}, \ref{QCN=3}, \ref{QCN=4}) for carrying out an imaginary time step can be decomposed into 4 parts, the state preparation sub-circuits $SP(N)$ and $SP(2N)$, the ITE operator $\hat{O}_P$, the contribution combiner $\hat{O}_H$, and the measurement sub-circuits.

$\hat{O}_P$ consists of $3N$ doubly-controlled Pauli gates (CCX, CCY, and CCZ), each of which contains five CNOT gates in their respective decomposition \cite{PhysRevA.52.3457}. Hence $\hat{O}_P$ contributes $O\left(N\right)$ CNOT gates to the circuit, while previous work implementing non-unitary operators using one ancillary qubit requires from $O\left(2^N\right)$\cite{Quantum.8.1311} to $O\left(4^N\right)$\cite{arXiv:quant-ph/0504100} CNOT gates. $\hat{O}_H$ and measurement sub-circuits consist only of the 1-qubit Hadamard gate $H$ and do not add any CNOT gate to the circuit.

On the other hand, the extra ancillary qubits with our method compared to some other methods comes with the disadvantage of a higher rate of discounted measurements when the ancilla qubits are not all 0.  Though this can be mitigated by increasing the shot count, it should be taken into account along with e.g. circuit depth when choosing between algorithms.

The state preparation sub-circuit $SP(2N)$ is where the bottleneck of the algorithm lies. In our algorithm, we adopt a simple but inefficient way to perform quantum state preparation. In this implementation, the number of CNOT gates scales as $O\left(2^{2N}\right)$. This is acceptable in the small basis expansion we showed. In general this amount of CNOT gates is required, as it corresponds to the $2^{2N+1}-2$ independent coefficients of the $2N$-qubit statevector \cite{PhysRevA.83.032302}. There has been attempts to prepare states more efficiently via different approaches \cite{IEEETransComput-AidedDesIntegrCircuitsSyst.43.161, PhysRevLett.129.230504, arXiv:2306.16831}. A recent work using tensor networks proved to achieve linear efficiency for up to 250 qubits \cite{QuantumSciTechnol.8.035027}. With improved state preparation algorithms, our work will prove to be an exponential improvement of its classical counterpart.

\section{Conclusion}
We have presented an implementation of the quantum imaginary time evolution method to solve the Hartree-Fock equations in the case of the helium-4 nucleus.  The method is demonstrated to be equivalent to the classical imaginary time evolution algorithm, with a resource scaling superior to the classical case.

We point out that the method will only work where the desired wave function, as represented in our work in oscillator expansion coefficients, is real.  This should cover many problems of interest, but e.g. seeking generally time-dependent states, or those resulting from correlated HF states would need a further algorithm to build upon the states found in our work.  Indeed, the Hartree-Fock technique finds approximate solutions of the nuclear many-body problem that are nearly uncorrelated (except for Pauli exclusion), while much of the promise of quantum computing lies in its ability to simulate highly-correlated states.  Hence, the utility of the quantum computation method here is limited to the scaling of the Hilbert space size with the number of qubits.  On the other hand, the Hartree-Fock solution is the standard starting point for building correlated wave functions through e.g. density-matrix methods \cite{PhysRevC.103.064304}, or as basis for shell model calculations where classically-obtained HF calcualtions have been used to start a quantum algorithm \cite{PhysRevC.105.064308}. Our algorithm can thus serve as the starting point for studies in which the ability of quantum computers to efficiently access highly correlated states is exploited.  Work along these lines is under investigation. 

\section*{Acknowledgements}
We acknowledge funding from the UK Science and Technology Facilities Council (STFC) under grant numbers ST/V001108/1 and ST/W006472/1, and from SEPNET.

\bibliography{MyLibrary}

\end{document}